\PassOptionsToPackage{hyphens}{url}
\documentclass[twocolumn,prl,amssymb,noeprint,superscriptaddress,twocolumn,longbibliography]{revtex4-2}

\usepackage{graphicx}
\usepackage{amsfonts}
\usepackage{amsmath}
\usepackage[usenames,dvipsnames]{color}
\usepackage{bm}
\usepackage{amssymb}
\usepackage{braket}
\usepackage{times}
\usepackage{siunitx}
\usepackage{multirow}
\usepackage{epstopdf}
\usepackage[normalem]{ulem}
\usepackage{hyperref}
\hypersetup{
colorlinks=true,
linkcolor=blue,
citecolor=blue,
filecolor=green,
urlcolor=blue,
}
\usepackage[nameinlink,capitalise]{cleveref}
\begin{document}

\title{Magnetic Feshbach resonances in Ba$^+$+Li collisions due to strong spin-orbit coupling
}

\author{Masato Morita}
\thanks{These authors contributed equally to this work.\\ \textcolor{blue}{masato.morita@fuw.edu.pl}}

\affiliation{Faculty of Physics, University of Warsaw, Pasteura 5, 02-093 Warsaw, Poland}

\author{Joachim Siemund}
\thanks{These authors contributed equally to this work.\\ \textcolor{blue}{masato.morita@fuw.edu.pl}}
\affiliation{Physikalisches Institut, Albert-Ludwigs-Universit\"at Freiburg, Hermann-Herder Str. 3, 79104  Freiburg, Germany}

\author{Wei Wu}
\affiliation{Physikalisches Institut, Albert-Ludwigs-Universit\"at Freiburg, Hermann-Herder Str. 3, 79104  Freiburg, Germany}

\author{Daniel von Schoenfeld}
\affiliation{Physikalisches Institut, Albert-Ludwigs-Universit\"at Freiburg, Hermann-Herder Str. 3, 79104  Freiburg, Germany}

\author{Jonathan Grieshaber}
\affiliation{Physikalisches Institut, Albert-Ludwigs-Universit\"at Freiburg, Hermann-Herder Str. 3, 79104  Freiburg, Germany}
    
\author{Agata Wojciechowska}
\affiliation{Faculty of Physics, University of Warsaw, Pasteura 5, 02-093 Warsaw, Poland}

\author{Krzysztof Jachymski}
\affiliation{Faculty of Physics, University of Warsaw, Pasteura 5, 02-093 Warsaw, Poland}

\author{Thomas Walker}
\affiliation{Physikalisches Institut, Albert-Ludwigs-Universit\"at Freiburg, Hermann-Herder Str. 3, 79104  Freiburg, Germany}
    \affiliation{Blackett Laboratory, Imperial College London, Prince Consort Road, London SW7 2AZ, United
    Kingdom}

\author{Fabian Thielemann}
\affiliation{Physikalisches Institut, Albert-Ludwigs-Universit\"at Freiburg, Hermann-Herder Str. 3, 79104  Freiburg, Germany}
\affiliation{5. Physikalisches Institut and Center for Integrated Quantum Science and Technology,
Universität Stuttgart, Pfaffenwaldring 57, 70569 Stuttgart, Germany}

\author{Tobias Schaetz}
\email{tobias.schaetz@physik.uni-freiburg.de}
\affiliation{Physikalisches Institut, Albert-Ludwigs-Universit\"at Freiburg, Hermann-Herder Str. 3, 79104  Freiburg, Germany}

\author{Micha{\l} Tomza}
\email{michal.tomza@fuw.edu.pl}
\affiliation{Faculty of Physics, University of Warsaw, Pasteura 5, 02-093 Warsaw, Poland}

\begin{abstract}
We report a pronounced dependence of magnetic Feshbach resonance spectra on the initial hyperfine-Zeeman state of Li in ultracold $^{138}$Ba$^+$+$^6$Li collisions. The measured number and distribution of resonances differ significantly between the two lowest states despite their similar electron spin character. We address this puzzle by developing a comprehensive yet generic computational model calibrated against key statistical features in the experimental spectrum. We confirm that strong spin-orbit coupling induces essential changes in the distribution of resonances, leading to an increase in the number of resolvable resonances. Our model reproduces the statistics of the spectrum with the lowest Li state but struggles with the second-lowest state, where a significantly smaller number of resonances is experimentally observed.

\end{abstract}
\maketitle

Recent advances in technologies for cooling and trapping ions, atoms, and molecules offer unique opportunities to explore cold and ultracold collisions and chemistry in unprecedentedly controlled environments with precisely prepared initial states \cite{BohnScience17,LangenNatPhys24,ChinRMP10,HarterContemPhys14,TomzaRMP19,DeisNatPhys24,KarmanNatPhys24}. 
These developments are central to emerging platforms for quantum information processing \cite{BruzewiczAPR19} and architectures for quantum simulations \cite{FossFeigAnnuRevCondMattPhys24,GeorgescuRMP14,AltmanPRXQuantum21}. 
In particular, hybrid approaches combining ions and neutral atoms open promising new research directions, by leveraging the high controllability at the single-ion level and the high density of trapped neutral atoms \cite{HarterContemPhys14,TomzaRMP19}. Successful demonstrations of control over ion-atom collision outcomes via initial state preparation for reactants \cite{RatschbacherNatPhys12,HallMolPhys13,SikorskyNatComm18,FeldkerPRA18,KatzNatPhys22} and more recent observations of magnetically tunable Feshbach resonances \cite{WeckesserNature21,ThielemannPRX25} unveil the potential of ion-atom systems as platforms for studying quantum information and ultracold chemistry.

Interatomic interactions in an ion-atom system are substantially different from those in the neutral atom-atom system with the same internal structure, due to the relatively long-range nature ($\propto -1/R^{4}$) of the polarization interaction potential \cite{TomzaRMP19}.
The resulting higher density of near-threshold bound states in the ion-atom system facilitates access to more Feshbach resonances \cite{TomzaPRA15}. However, the higher density makes the assignment of the resonances challenging. In addition, uncertainties in interaction potentials and the sensitivity of observables at ultracold temperatures to tiny changes in the potentials practically prohibit the \textit{a priori} theoretical prediction of individual resonances \cite{WallisEPJD11,FurstPRA18,MoritaPRL19,SikorskyPRL18}, even though converged scattering calculations are feasible. This necessitates a synergistic approach combining state-of-the-art experiments and calculations to mitigate these difficulties~\cite{WalewskiSA25}.

In a previous study \cite{WeckesserNature21},
we performed ion-loss spectroscopy measurements for a single $^{138}$Ba$^+$($^2$S) ion immersed in a gas of $^6$Li($^2$S) atoms prepared in the second lowest energy state $\ket{2}_\mathrm{Li}$ in several magnetic field ($B$-field) regions.
Some of the observed resonances have been preliminarily assigned using the asymptotic bound-state model (ABM) \cite{TieckePRA10}, highlighting the necessity of including the higher-partial waves and the second-order spin-orbit (SO) coupling effect to explain the observed spectrum.

More recently, the loss spectrum of $^{138}$Ba$^+$ in a gas of $^{6}$Li atoms prepared in the lowest state $\ket{1}_\mathrm{Li}$ was measured at lower temperatures than those used in previous studies \cite{ThielemannPRX25}. 
Remarkably, compared to the spectra recorded in the earlier work \cite{WeckesserNature21}, we found that $\ket{1}_\mathrm{Li}$ exhibits markedly more resonances than $\ket{2}_\mathrm{Li}$, although their energies show similar magnetic-field dependence due to comparable electron spin character. This astonishing observation points to richer underlying physics than previously anticipated and calls for a systematic study.

In the present Letter, we address this complexity by analyzing the number and distribution of magnetic Feshbach resonances for the collisions between Ba$^+$ and Li by combining systematic experimental measurements and scattering calculations. Loss spectra for a single $^{138}$Ba$^+$ are observed with two distinct initial states of $^{6}$Li, $\ket{1}_\mathrm{Li}$ and $\ket{2}_\mathrm{Li}$ in a common range of the magnetic field $B=240-340$~G, where the electron spin in both these states is well characterized as the pure down spin state ($M_{S_\mathrm{Li}}=-1/2$) due to the dominance of the electron spin Zeeman interaction over the hyperfine coupling. 
The Schr\"odinger equation for the two-body collision, including the SO coupling effect that was not previously treated in detail, is rigorously solved using the coupled-channel (CC) method \cite{ArthursPRS60,TscherbulPRL16,FurstPRA18}. 
To compensate for the uncertainty in the interaction potentials, we exploit a statistical approach to identify the optimal potentials by assessing the fidelity of resonance positions between the experiment and the calculation. 
This enables a consistent reproduction of the observed statistical properties of the spectrum, providing a transferable framework for other systems.

\begin{figure}[t!]
\centering
\includegraphics[width=0.95\columnwidth]{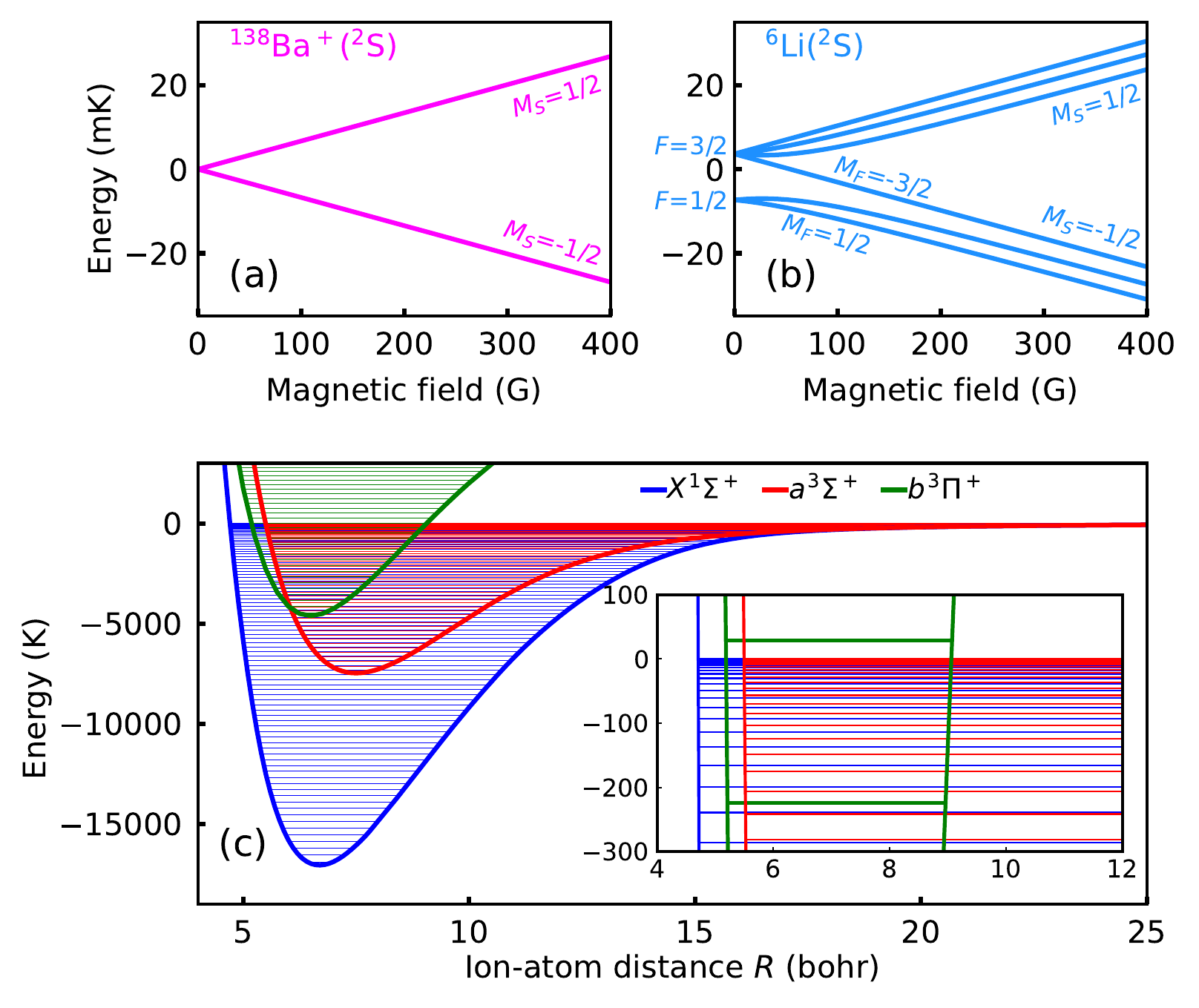}
    \vspace*{-0.3cm}
    \caption{Energy levels of hyperfine-Zeeman states (a) $^{138}$Ba$^+$($^2$S) and (b)$^6$Li($^2$S) as functions of magnetic field. 
    (c) {\it ab initio} potential energy curves for the singlet $X ^1\Sigma^+$ (blue) and the triplet $a ^3\Sigma^+$ (red) states, dissociating into Ba$^+$($^2$S) and Li($^2$S). At an ion-atom distance of $R=6$~bohr, $b^3\Pi$ (green) crosses and strongly couples with $a ^3\Sigma^+$ by the spin-orbit (SO) interaction. Bound vibrational levels are displayed via horizontal lines. The inset highlights the levels of $b^3\Pi$ (green) near the collision threshold.}
\label{Fig_ChE}
\end{figure}

In our experiment, we employ ion-loss spectroscopy to reveal the Feshbach resonance spectrum. Full details are provided in Refs.~\cite{WeckesserNature21,ThielemannPRX25}. We provide a short summary in the following. In our hybrid setup we use a linear Paul trap to confine and prepare a single, Doppler-cooled $^{138}$Ba$^+$ ion. We compensate stray electric fields to within $E_\text{stray}\approx 3$~{mV/m}. 
It is inserted into a cloud of ultracold $^6$Li atoms ($n_\mathrm{Li}\approx 5\times10^{17}$/m$^{3}$) held in a crossed optical dipole trap (xODT) at 1064$\,$nm and spin-polarized in either the lowest hyperfine-Zeeman state $\ket{1}_\mathrm{Li}$ or the second lowest $\ket{2}_\mathrm{Li}$ (\cref{Fig_ChE} (b)). 
After optical pumping to the S$_{1/2}$ electronic ground state, the Ba$^+$ ion remains in an unpolarized mixture of $S_\mathrm{Ba^+}=1/2$.
The ion interacts with the atomic bath for $t_\text{int}=200$~{ms} at a given magnetic field $B$, allowing the ion spin relaxation into the lowest Zeeman state (\cref{Fig_ChE} (a)). Three-body recombination can lead to heating the ion at Feshbach resonance positions.
The ion's product state is measured to reconstruct its survival probability $P_\text{surv}$. 
The bath temperature of Li is controlled via the depth of xODT and thus depends on the xODT light intensity during interactions. 
A near-resonant light pulse is used to deplete the number of atoms to maintain a constant $n_\mathrm{Li}$ across different $T_\mathrm{Li}$.

\begin{figure}[t!]
 \centering
    \includegraphics[width=0.92\columnwidth]{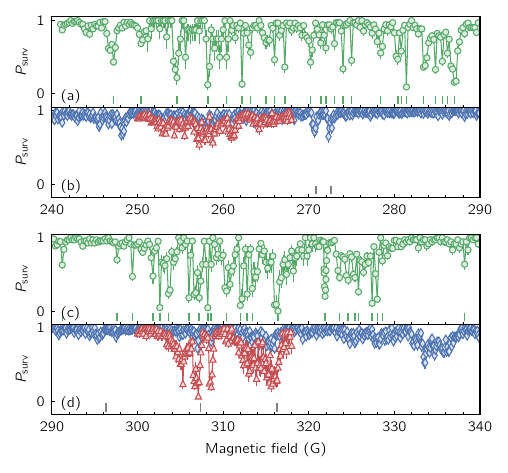}
    \vspace*{-0.1cm}
    \caption{Ion-loss spectra as a function of magnetic field for Li prepared in the lowest $\ket{1}_\mathrm{Li}$ [(a) and (c), green circles] and second lowest $\ket{2}_\mathrm{Li}$ [(b) and (d), blue diamonds and red triangles] hyperfine-Zeeman states. Significant features (green dashes) in the former spectrum are identified based on the procedure described in the main text. We also mark the previously reported resonance positions (gray dashes) \cite{WeckesserNature21}. 
    The bath temperature is $T_\mathrm{Li}\approx 700$~nK (green circles and blue diamonds), and we scanned over a limited range 
    with an increased temperature of $T_\mathrm{Li}\approx 1.4~\mu$K for $\ket{2}_\mathrm{Li}$
    (red triangles) \cite{expfootnote}.
All measurements are performed at a Li density of $n_\mathrm{Li} \approx 5 \times 10^{17}$/m$^{3}$.
The error bars depict 1$\sigma$ confidence intervals.  
    }
    \label{Fig_EXP}
\end{figure}

\Cref{Fig_EXP} shows the resulting loss spectra for Li in $\ket{1}_\mathrm{Li}$ and $\ket{2}_\mathrm{Li}$ recorded with resolutions of $\Delta B= 200$~{mG} and {100}~{mG}, respectively. Part of the $\ket{1}_\mathrm{Li}$ spectrum has been published in Ref.~\cite{ThielemannPRX25}. 
Based on a significance of at least 3$\sigma$, relative to neighboring features, we identify 49 resonances for $\ket{1}_\mathrm{Li}$ at $T_\mathrm{Li}\approx 700$~{nK}. 
In contrast, the $\ket{2}_\mathrm{Li}$ spectrum at the same temperature exhibits fewer and significantly weaker features. Increasing the bath temperature to $T_\mathrm{Li}\approx 1.4~\mu$K  enhances the signal strength, consistent with the spectrum for $\ket{1}_\mathrm{Li}$ and in agreement with previous findings \cite{WeckesserNature21}. 
However, even at the higher temperatures, the density of resonances is significantly lower than that observed for $\ket{1}_\mathrm{Li}$.

Here, we summarize our calculations for collisions between Ba$^+$($^2$S) and Li($^2$S) in an external $B$-field.  
The Hamiltonian is given as (in atomic units)  
\small
\vspace{-0.1cm}
\begin{equation}
\hat{H} = - \frac{1}{2\mu R}  \frac{\partial^2}{\partial R^2} R 
+ \frac{\boldsymbol{\hat{l}}^2}{2\mu R^2} 
+ \hat{H}_{\mathrm{Ba^+}}
+ \hat{H}_{\mathrm{Li}}
+ \hat{H}_{\mathrm{int}},
\label{eq:Heff}
\end{equation}
\normalsize
where $\mu$ is the reduced mass, $R$ is the distance between Ba$^+$ and Li, $\boldsymbol{\hat{l}}$ denotes the orbital angular momentum operator for their relative motion. $\hat{H}_{\text{Ba}^+}$ and $\hat{H}_\text{Li}$ indicate the Hamiltonians for the isolated Ba$^+$ and Li in an external $B$-field, dictating their internal hyperfine-Zeeman states as shown in \cref{Fig_ChE}~(a) and (b). The nuclear spin is zero for $^{138}$Ba$^+$ while the nuclear spin of $^{6}$Li, $I_\mathrm{Li}=1$, couples with the electron spin via the hyperfine interaction. 
However, in the present $B$-field region, the Zeeman interaction for the electron spin is strong enough to decouple the hyperfine coupling. 
Consequently, $\ket{1}_\mathrm{Li}$ and $\ket{2}_\mathrm{Li}$ are expressed with a common electron spin character as $\ket{M_{S_\mathrm{Li}}=-1/2}\ket{M_{I_\mathrm{Li}}=1}$ and $\ket{M_{S_\mathrm{Li}}=-1/2}\ket{M_{I_\mathrm{Li}}=0}$, respectively, giving rise to a quite similar $B$-field dependence of their energies (\cref{Fig_ChE}).
The last term $\hat{H}_{\mathrm{int}}$ is given as the sum of the electrostatic interaction $\hat{V}(R)$, the magnetic dipole-dipole interaction $\hat{V}_\mathrm{dd}(R)$, and the second-order spin-orbit interaction $\hat{V}_\mathrm{SO}(R)$.

$\hat{V}(R)$ is expressed with the potentials, $X^1\Sigma^+$ and $a^3\Sigma^+$ in \cref{Fig_ChE}~(c), as $\hat{V}(R)=\sum^{}_{S, M_S} |S M_S \rangle V_S(R) \langle S M_S|$, where $S$ and $M_S$ denote the total electron spin quantum number and its projection of the collision complex, and $V_S(R)$ denotes the singlet ($S=0$) and triplet ($S=1$) potentials.  
$\hat{V}(R)$ conserves $S$ and $l$ and their projections $M_S$ and $M_l$ while the difference between singlet and triple potentials in the short-range may cause the electron spin-exchange between Ba$^+$ and Li. 
In contrast, $\hat{V}_\mathrm{dd}(R)$ and $\hat{V}_\mathrm{SO}(R)$ introduce the couplings that may change $l$ as well as $S$, and is given as
\cite{MiesJRNIST96,TscherbulPRL16,FurstPRA18,SardarPRA23} 
\vspace{-0.1cm}
\small
\begin{equation}
\sqrt{\frac{24\pi}{5}} \left[-\frac{\alpha ^2}{R^3}+\lambda_\mathrm{SO}(R)\right]  \sum^{}_{q} (-1)^q Y_{2,-q}^*(\hat{\bm{R}}) [\hat{\bm{S}}_\mathrm{Ba^+}\otimes \hat{\bm{S}}_\mathrm{Li}]_{q}^{(2)},
\label{eq:H_dipolar}
\end{equation}
\normalsize
where $\alpha$ is the fine-structure constant 
and $[\hat{\bm{S}}_\mathrm{Ba^+}\otimes \hat{\bm{S}}_\mathrm{Li}]_{q}^{(2)}$ denotes a spherical tensor product of electron spins. The SO parameter $\lambda_\mathrm{SO}(R)$ reflects an effective short-range correction to $\hat{V}_\mathrm{dd}(R)$, arising from the SO interaction between $^3\Sigma^+$ and $^3\Pi$~\cite{TscherbulPRL16,TomzaRMP19,WeckesserNature21, SardarPRA23, AkkariPhysScr24, XingJPB24}. However, the crossing of $^3\Sigma^+$ and $^3\Pi$ may significantly enhance or suppress the effective coupling strength, depending on the exact energy distances of vibrational levels in $^3\Pi$ from the collision threshold [see the inset in \cref{Fig_ChE}(c)]. We note that $\hat{V}_\mathrm{SO}(R)$ and $\hat{V}_\mathrm{dd}(R)$ lift the degeneracy of partial waves and increase the number of Feshbach resonances.
Although $\hat{H}_\mathrm{int}$ commutes with the square of the total angular momentum operator $\hat{\bm{J}}_\mathrm{tot}^2$ for the collision complex, $J_\mathrm{tot}$ is not conserved in the $B$-field due to the Zeeman interactions, thus only the total angular momentum projection  $M_\mathrm{tot} = M_{S_{\mathrm{Ba^+}}}+M_{S_{\mathrm{Li}}}+M_{I_{\mathrm{Li}}}+M_l$ and the parity $p=(-1)^l$ are conserved.

The time-independent Schr\"odinger equation is solved with the coupled-channel (CC) method \cite{ArthursPRS60,TomzaRMP19}, expanding the wavefunction using a basis set for the angular momenta as $\ket{\Psi}=\sum_j \Phi_j(R)\ket{\Theta_j} /R$, where the basis sate $\ket{\Theta_j}$ is 
$\ket{\Theta_j} =\ket{S_\mathrm{Ba^+}\, M_{S_\mathrm{Ba^+}}} \ket{S_\mathrm{Li}\, M_{S_\mathrm{Li}}} \ket{I_\mathrm{Li}\, M_{I_\mathrm{Li}}} \ket{l\,M_{l}}$.
The CC equation to be satisfied by the expansion coefficients $\{\Phi_{j}\}$ is
\small
\vspace{-0.2cm}
\begin{equation}
\begin{split}
\left[\frac{d^2}{dR^2}\right. & \left.-\frac{l(l+1)}{R^2} +2\mu E \right]{\Phi}_j(R)  \\ & =
        2\mu \sum^{}_{k} \langle \Theta_j | \hat{H}_{\mathrm{Ba^+}} + \hat{H}_{\mathrm{Li}} + \hat{H}_{\mathrm{int}} |\Theta_{k}\rangle {\Phi}_{k}(R).
\label{eq:CC}
\end{split}
\end{equation}
\normalsize
This can be decomposed into smaller independent CC equations specified by $M_\mathrm{tot}$ and $p$ of the basis states $\ket{\Theta_j}$.

In this Letter, we first calculate the elastic cross section $\sigma^\mathrm{el}(E_\mathrm{c})$, where $E_\mathrm{c}$ is the collision energy, from the S-matrix obtained by solving the CC equations with the log-derivative propagation method and the subsequent matching to the scattering boundary condition \cite{JohnsonJCP77,ManolopoulosJCP86}. 
We then calculate the thermally averaged rate constant $K^\mathrm{el}(T)$ to account for resonance overlap caused by profile broadening or narrowing due to couplings and thermal effects. We calculate $K^\mathrm{el}$ by taking a convolution of $\sigma^\mathrm{el}(E_\mathrm{c})$ with the Maxwell-Boltzmann distribution at the temperature of $T=0.8~\mu$K, based on our previous simulation \cite{ThielemannPRX25} (see also End Matter).

The unambiguous assignment of magnetic Feshbach resonances in ultracold collisions is a significant challenge for current theoretical studies \cite{WallisEPJD11,FurstPRA18,MoritaPRL19,SikorskyPRL18} due to the uncertainties in the interaction potentials and the extreme sensitivity of the scattering observables to tiny variations in the potentials.
Therefore, rather than attempting direct assignments for individual resonances, we introduce a generic approach to optimize the potentials by comparing the statistical distribution of resonances in the calculated $K^\mathrm{el}$ and the experimental loss spectrum.  
We then assess the validity of our approach by analyzing the statistical properties of resonances in $K^\mathrm{el}$. 
Finally, we explore the initial Li state dependence using the optimized potentials.
We note that $K^\mathrm{el}$ and the experimental loss spectrum are different properties. Therefore, it is not possible to directly compare the magnitudes and the resonance profiles between these spectra. 
Our primary purpose in calculating $K^\mathrm{el}$ is to avoid overcounting of detectable resonances, arising from neglecting their widths, magnitudes, or overlaps with other resonances in a collision energy distribution.

We begin our optimization by determining a search range for the singlet and triplet potentials to prevent unbounded growth of the search space.
We generate a sample of potentials by scaling the {\it ab initio} potentials with scaling factors \cite{WallisEPJD11,MoritaPRL19} as $\lambda_0 V^\mathrm{ref}_{0}(R)$ and $\lambda_1 V^\mathrm{ref}_{1}(R)$, where $V^\mathrm{ref}_{0}(R)$ and $V^\mathrm{ref}_{1}(R)$ are {\it ab initio} potentials for $X^1\Sigma$ and $a^3\Sigma$. 
We determine the ranges of the scaling factors as $0.9816\le \lambda_0 \le 1.0027$ and $0.9789 \le \lambda_1 \le 1.0094$ based on the peak-to-peak region of the scattering length $a_S$ on each potential, including $\lambda_S=1$ (see End Matter). 
This selection covers the full range of possible values of scattering phase shift, \{$-\pi/2,\pi/2$\}, for each potential, and ensures a one-to-one correspondence between $\lambda_S$ and $a_S$. The scaled potentials in the sample are specified by the values of ($a_0$, $a_1$) as well as ($\lambda_0, \lambda_1$).

Our purpose is to identify the potentials leading to $K^\mathrm{el}$ that captures statistical characteristics of the experimental spectra. 
As shown in \cref{Fig_EXP}, the experimental spectrum with $\ket{1}_\mathrm{Li}$ exhibits a sufficiently large number of resonances, allowing meaningful statistical analysis. We calculate $K^\mathrm{el}$ with $\ket{1}_\mathrm{Li}$ using each pair of scaled singlet and triplet potentials in the sample and compare the resonance distributions with the experimental spectrum.
To this end, we introduce an objective function
$D^2$ to quantify the similarity, defined as the sum of the squares of the minimum distances between each experimental resonance and the closest theoretical resonance,
\small
\begin{equation}
D^2(a_0, a_1)=\sum_{i}^{} \underset{j}{\mathrm{min}}(B^{\mathrm{exp}}_i-B^{\mathrm{CC}}_j(a_0,a_1))^2,
\label{eq:chi}
\end{equation}
\normalsize
where $B^{\mathrm{exp}}_i$ denotes the $i$-th resonance position in the experimental spectrum, $B^{\mathrm{CC}}_j(a_0,a_1)$ is the $j$-th resonance position in $K^\mathrm{el}$ obtained with the scaled potentials specified by ($a_0$, $a_1$). 
We identify the potentials, resulting in the minimum value of $D^2$ as 91.2, with the scaling parameters 
 ($\lambda_0$, $\lambda_1$)=(0.9816, 0.9789) corresponding to ($a_0$, $a_1$)=($7.152\times 10^5$, $1.838\times 10^5$) bohr. Hereafter, we refer to these potentials as the {\it 
optimal} potentials (see End Matter). 
We observe 41 resonances in $B=240-340$~G with the optimal potentials, comparable to 49 resonances in the experimental spectrum. 
In our calculations, Ba$^+$ is assumed to be initially in the lowest Zeeman state (\cref{Fig_ChE}(a)) by the spin-relaxation due to the SO interaction.

$D^2$ evaluates a fidelity of resonance positions, not the total number of resonances or the correlations between their positions. 
Nevertheless, we obtained a reasonable result regarding the number of resonances, which encourages us to investigate other statistical properties to further support the suitability of the optimal potentials.
We first calculate the staircase function \cite{BrodyRMP81}, counting the number of resonances up to a given magnetic field $B$ as
\vspace{0.0cm}
\small
\begingroup
\setlength\abovedisplayskip{6pt}
\setlength\belowdisplayskip{6pt}
\begin{equation}
\mathcal{N}(B)=\int_{-\infty}^{B} dB^\prime \sum_{i}^{} \delta(B^\prime-B_i),
\label{eq:stair}
\end{equation}
\normalsize
where $B_i$ denotes the $i$-th resonance position in $B=240-340$~G.
In \cref{Fig_PDF} (a), $\mathcal{N}(B)$ for the experiment spectrum exhibits a large slope compared to those obtained with the scaled potentials at high magnetic fields and takes the largest value of $\mathcal{N}(B)$ at $B=340$~G.  
We observe that the optimal potentials provide one of the best fits to the experimental $\mathcal{N}(B)$. 

\begin{figure}[t!]
 \centering
\includegraphics[width=0.85\columnwidth]{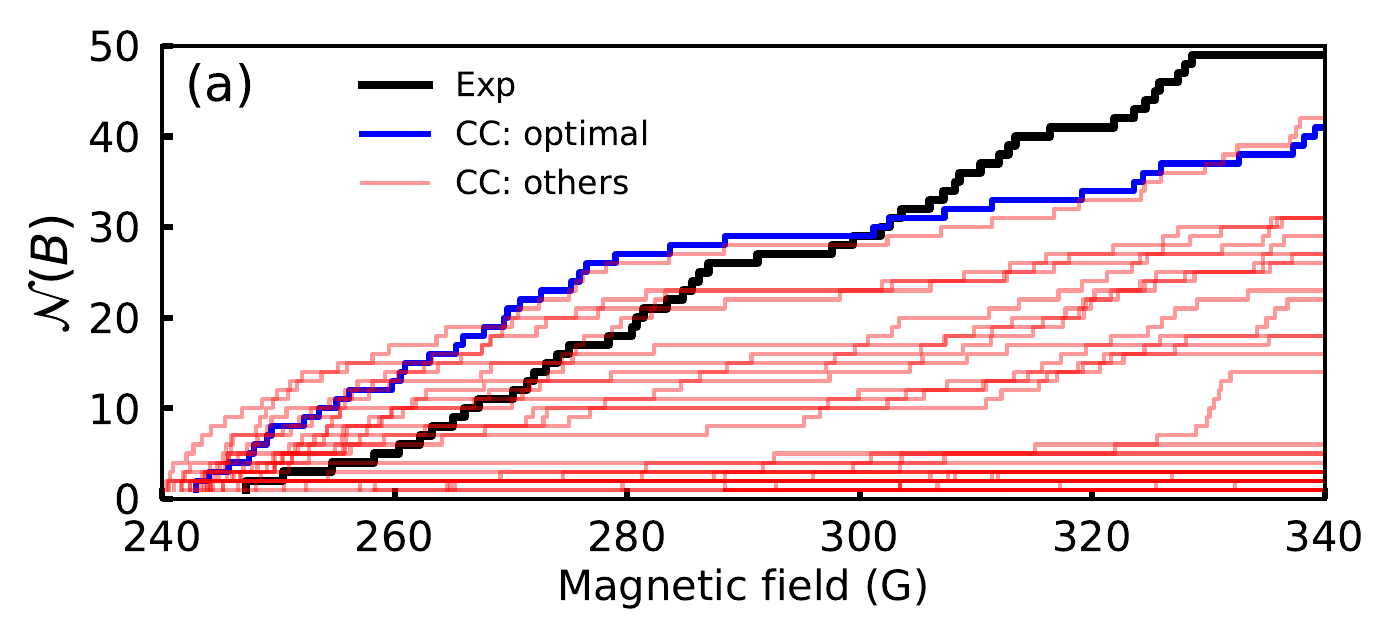}
\includegraphics[width=1.0\columnwidth]{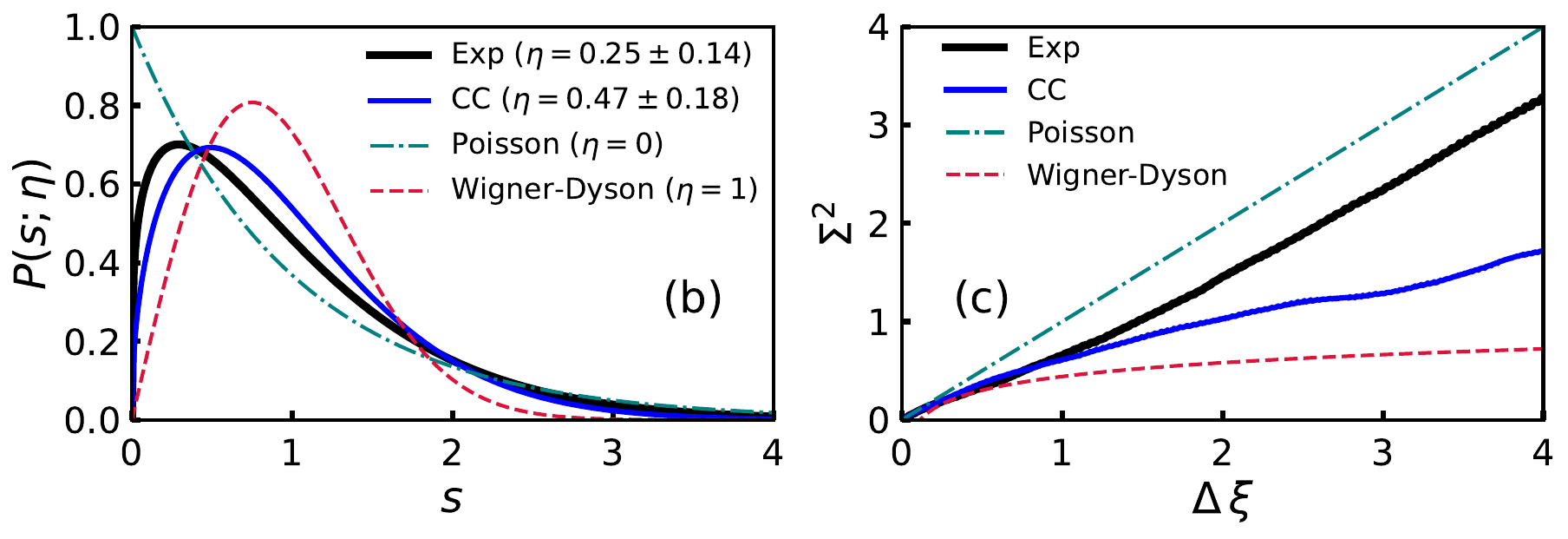}
    \vspace*{-0.6cm}
    \caption{Distribution of resonances for the $\ket{1}_\mathrm{Li}$ initial state. (a)
     Staircase functions  $\mathcal{N}(B)$ show the numbers of resonances from $B=240$~G up to a magnetic field $B$. 
     (b) Probability density of unfolded nearest neighbor spacing $s$ expressed using the Brody distribution, (black) experiment and (blue) calculation with the optimal potentials. 
     For reference, Poissonian exponential ($\eta=0$) and chaotic Wigner-Dyson ($\eta=1$) distributions are shown with dashed curves.  
    (c) Number variance as a function of the interval length $\Delta\xi$. 
    }
    \label{Fig_PDF}
\end{figure}

The correlations of resonance positions are examined based on the Gaussian orthogonal ensemble (GOE) \cite{BrodyRMP81,GuhrPR98,WeidenmullerRMP09} in the standard random matrix theory (RMT) \cite{FrischNature14,FryePRA16,GreenPRA16,KosickiNJP20}.
The nearest neighbor spacing (NNS) is the most widely used property to diagnose possible quantum chaos in a distribution of resonances. The staircase functions $\mathcal{N}(B)$ in \cref{Fig_PDF} (a) do not behave as linear curves, indicating the presence of systematic variations of the densities of resonance with $B$-field. 
Unfolding has been used as a practical way to eliminate such variations that lead to differences in the spacing scale \cite{FryePRA16,GreenPRA16,KosickiNJP20}. We fit a 4th-order polynomial to $\mathcal{N}(B)$ to obtain a smooth function $\xi(B)$. We then map the sequence of resonance positions \{$B_i$\} ($i=1,2,..$) onto the dimensionless numbers \{$\xi_i$\} as $\xi_i=\xi(B_i)$. The resultant average density of unfolded \{$\xi_i$\} is unity, implying a unit mean NNS, where the NNS for the $i+1$ and $i$-th resonances is given as $s_i=\xi_{i+1}-\xi_i$  \cite{GuhrPR98, WeidenmullerRMP09,FryePRA16,GreenPRA16,KosickiNJP20}. 
This allows a direct comparison with the results of RMT. 
\Cref{Fig_PDF} (b) shows probability density for the distribution of unfolded NNS,
described by the Brody distribution
\cite{BrodyRMP81}
\small
\begin{equation}
P_\mathrm{B}(s;\eta)=c_{\eta}(\eta+1)s^{\eta}\, \mathrm{exp}(-c_\eta s^{\eta+1}), 
\label{eq:Brody}
\end{equation}
\normalsize
where the Brody parameter $\eta\, (0 \leq \eta \leq 1)$ is determined by the maximum likelihood estimation with \{$s_i$\} \cite{FryePRA16,GreenPRA16,KosickiNJP20} (see End Matter).
$c_{\eta}$ in \cref{eq:Brody} is expressed with the Gamma function as $c_{\eta} = [\, \Gamma( (\eta+2)/(\eta+1))\,]^{\eta+1}$.

In both the experiment (black) and the calculation with the optimal potentials (blue), the probability densities deviate from the pure chaotic Wigner-Dyson ($\eta=1$) and the random Poissonian exponential ($\eta=0$) distributions while the values of $\eta$ are slightly closer to the Poissonian. 
Since the spectra are observed without resolving $M_\mathrm{tot}$ and parity $p$, the Poissonian distribution is natural as a result of the overlap of independent resonance distributions \cite{GuhrPR98}. The observed slight deviations from the pure exponential distribution indicate a moderate dominance of specific $M_\mathrm{tot}$ and parity components.

Lastly, we explore the number variance $\Sigma^2$ to probe correlation between resonances beyond nearest neighbors \cite{BrodyRMP81,GuhrPR98, FryePRA16,KosickiNJP20}. 
$\Sigma^2$ is defined as a function of the interval length $\Delta \xi$ 
\vspace{-0.1cm}
\small
\begin{equation}
\Sigma^2(\Delta \xi) = \langle\hat{N}^2(\Delta \xi)\rangle-\langle\hat{N}(\Delta \xi)\rangle^2, 
\label{eq:variance}
\end{equation}
\normalsize
where $\hat{N}(\Delta \xi)$ denotes the number of resonances in an interval $\Delta \xi$. $\hat{N}^2(\Delta \xi)$ is its squared value. 
$\langle...\rangle$ denotes averaging over intervals with a common length $\Delta \xi$.
Due to unfolding, $\langle \hat{N}(\Delta \xi)\rangle \approx \Delta \xi$.
\Cref{Fig_PDF} (c) shows agreement between the calculation result with the optimal potentials and the experimental result at small interval lengths. 
For $\Sigma^2$, we also observe deviations from the pure Poissonian and Wigner-Dyson characteristics with increasing deviations at longer intervals.

\begin{figure}[t!]
 \centering
 \includegraphics[width=0.82\columnwidth]{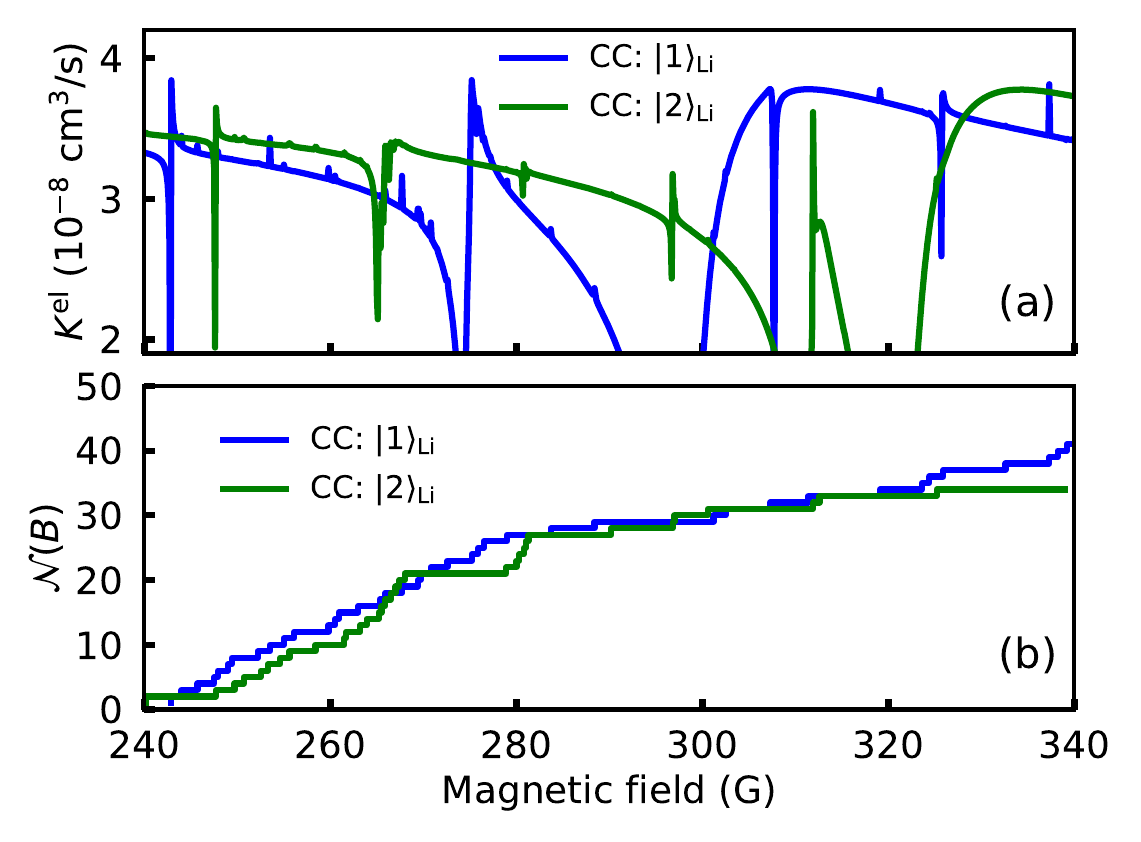}
    \vspace*{-0.4cm}
    \caption{ (a) Calculated thermally averaged rate constants $K^\mathrm{el}$ 
    with the lowest $\ket{1}_\mathrm{Li}$ and second lowest $\ket{2}_\mathrm{Li}$ initial states. 
    (b) Staircase functions for the resonances in the rate constants.  
    }
    \label{Fig_1vs2}
\end{figure}

Together, our statistical comparisons confirm that the optimal potentials not only reproduce key features of the resonance positions in $\ket{1}_\mathrm{Li}$, but also capture the correlations within the resonance distribution in a reasonable manner. We now extend our analysis to explore the observed state dependence in the spectra (\cref{Fig_EXP}).

To this end, we perform the CC calculations with $\ket{2}_\mathrm{Li}$ using the same optimal potentials. In \cref{Fig_1vs2} (a), we observe an evident initial state dependence for $K^\mathrm{el}$. However, the staircase functions $\mathcal{N}(B)$ in (b) indicate that the cumulative number of resonances is largely independent of the initial states. 
This aligns with the fact that both $\ket{1}_\mathrm{Li}$ and $\ket{2}_\mathrm{Li}$ are characterized by the common electron spin state of $\ket{M_{S_\mathrm{Li}}=-1/2}$, resulting in a similar $B$-field dependence for these states as shown in \cref{Fig_ChE} (b).
We also confirm a moderate positive correlation for the number of resonances between $\ket{1}_\mathrm{Li}$ and $\ket{2}_\mathrm{Li}$ (see End Matter). 
However, these results stand in sharp contrast to the experimental prediction in \cref{Fig_EXP}.

Notably, this unexpected initial state dependence in the experimental spectra could reflect an emergent mechanism or sensitivity to subtle environmental parameters not included in our present theory.
To identify the origin of the state dependence, it would be necessary to consider mechanisms beyond the two-body dynamics, such as secondary collisions, state-dependent three-body effects, time-dependent trapping potential~\cite{PinkasNatPhys23,HirzlerPRL23}, or unexpected measurement imperfections caused by laser, trap, or detection limits. 
Further exploration, such as extending the $B$-field range, using different initial Li states, or implementing spin-resolved detection, offers a promising path to resolve this puzzle. 

In summary, we experimentally observe remarkably distinct loss spectra for the two lowest hyperfine-Zeeman states of Li, differing in the number and strength of resonances.
This startling result is beyond our intuitive understanding because the energies and spin compositions of these states exhibit similar magnetic field dependences. 
In an effort to understand this difference, we performed rigorous coupled-channel calculations including spin-orbit interaction as well as collision energy distribution.
We propose an optimization approach that offers an effective framework for modeling interaction potentials, capturing key statistical features of magnetic Feshbach resonances. The remaining discrepancy between experiment and calculation reported in this Letter may indicate the relevance of more complex few-body and trap effects and offer a challenge to deepen our understanding of the lifetime associated with magnetic Feshbach resonances. Future investigations could benefit from employing purely optical trapping of both atoms and ions~\cite{SchmidtPRL20,SchaetzJPB17}, eliminating rf-induced disturbances. Experimental determination of the partial wave character of the resonances~\cite{ThielemannPRX25} would also offer deeper insight into the nature of the observed resonances. We will further investigate the origin of the discrepancy in our future work.  

\vspace{0.09cm}
\begin{acknowledgments}
{\it Acknowledgments --}
M.M thanks Maks Z. Walewski and Matthew D. Frye for stimulating discussions. We gratefully acknowledge the the European Union (ERC, 101042989 -- QuantMol) for financial support and the Poland’s high-performance computing infrastructure PLGrid (HPC Center: ACK Cyfronet AGH) for providing computer facilities and support (computational grant no.~PLG/2024/017844). This project has received funding from the European Research Council (ERC) under the European Union’s Horizon 2020 research and innovation program (Grant No. 648330), the Deutsche Forschungsgemeinschaft (DFG, Grant No. SCHA 973/9-1-3017959), and the Georg H. Endress Foundation. F.T., J.S., D.v.S., J.G. and T.S. acknowledge financial support from the DFG via the RTG DYNCAM 2717. W.W. acknowledges financial support from the QUSTEC programme, funded by the European Union’s Horizon 2020 research and innovation program under the Marie Sk{\l}odowska-Curie (Grant No. 847471). This work was supported by the Polish National
Agency for Academic Exchange (NAWA) via the Polish Returns 2019 programme.
\end{acknowledgments}

J.S., F.T., and J.G. conducted the experiments and analyzed the experimental data with support from T.W.. 
J.S., F.T., W.W., T.W. and D.v.S. maintained the experimental setup. 
T.S. and F.T conceived and supervised the experiments. 
M.T. conceived and supervised the theoretical calculations. M.M. established and conducted the scattering calculations and analyzed them with support from M.T.. 
M.M. wrote the first version of the manuscript with inputs from M.T., K.J., T.S, J.S., and F.T. 
All authors discussed the results and the final version of the manuscript.

\appendix*
\section{\large End Matter}

\vspace{-0.2cm}
{\it {\bf Ab initio potential energy curves} }

As described in our previous paper \cite{WeckesserNature21}, {\it ab initio} potential energy curves were calculated with the coupled cluster method restricted to single, double, and non-iterative triple excitations, CCSD(T), and the multireference configuration interaction method restricted to single and double excitations, icMRCISD, with aug-cc-pwCV5Z basis sets and ECP46MDF small-core pseudopotential for Ba$^+$. 
The MOLPRO \cite{WernerWIRES12} was used to perform {\it ab initio} electronic structure calculations.

\vspace{0.1cm}
{\it {\bf Scattering calculation} }

We use the reduced mass of $\mu=5.763722$ amu. The Hamiltonian for isolated Li is composed of hyperfine coupling and Zeeman interactions as
$\hat{H}_\mathrm{Li}=A_\mathrm{Li} \hat{\bm{I}}_\mathrm{Li} \cdot \hat{\bm{S}}_\mathrm{Li}
-g_I \mu_\mathrm{N} B \hat{I}_{\mathrm{Li},z}
+g_S \mu_\mathrm{B} B \hat{S}_{\mathrm{Li},z}$,
where the hyperfine coupling constant is $A_\mathrm{Li}=7.301$~mK for $^6$Li ($^2$S).
The direction of the $B$-field is defined as the $z$-axis. 
$\mu_\mathrm{N}$ and $\mu_\mathrm{B}$ are the nuclear and bohr magnetons,  respectively. $g_I$ and $g_S$ are the $g$-factors for the nuclear spin of $^6$Li and for the electron.
The resultant hyperfine splitting between $F_\mathrm{Li}=3/2$ and $1/2$ states is $\Delta E=10.95$~mK. Due to the absence of nuclear spin, $\hat{H}_{\mathrm{Ba}^+}$ includes only the Zeeman interaction of the electron.  
The interaction potential energy curves for singlet $X ^1\Sigma^+$ and triplet $a ^3\Sigma^+$ converge to the same dissociation energy and behave as $-C_4/R^4$ \cite{TomzaRMP19} in the long-range. $C_4=80.91$ in atomic units.

For each $M_\mathrm{tot}$ and parity $p$ component, we numerically solve the CC equation with the MOLSCAT program suite \cite{HutsonCPC19_MOLSCAT,MOLSCAT}. 
The CC equations are integrated from $R_\mathrm{min}=4$~bohr to $R_\mathrm{mid}=30$~bohr in steps of $\Delta R=0.005$~bohr using the diabatic modified log-derivative propagator \cite{ManolopoulosJCP86}, and then propagate until $R_\mathrm{max}=10^4$~bohr using the variable step Airy propagator \cite{AlexanderJCP87}. 
Matching the solutions to the boundary conditions yields the $S$-matrix, from which we obtain the elastic cross section $\sigma^\mathrm{el}(E_\mathrm{c})$ as 
\small
\begingroup
\setlength\abovedisplayskip{1.5pt}
\setlength\belowdisplayskip{1.5pt}
\begin{equation}
\sigma^\mathrm{el}(E_\mathrm{c}) = \frac{\pi^2}{k} \sum_i|1-S_{ii}(E_\mathrm{c})|,
\label{eq:cross}
\end{equation}
\normalsize
where $E_\mathrm{c}$ and $k$ are the collision energy and the wavevector for the channel. $S_{ii}$ denotes the diagonal S-matrix element for a partial wave component. 
We employ the basis set containing partial waves up to $l_\mathrm{max}=5$ ($h$-wave), for $-3 \le M_\mathrm{tot} \le 3$. We scan $B$-field in the range of $B=240-340$~G with an interval of $\Delta B=0.1$~G.

Different from $\sigma^\mathrm{el}(E_\mathrm{c})$ given at each $E_\mathrm{c}$, the experimental spectra observed in the range $B=240-340$~G do not have such a fine resolution about $E_\mathrm{c}$. To simulate spectra by taking into account the collision energy distribution, we evaluate the thermally averaged rate constant $K^\mathrm{el}(T)$ at $T=0.8~\mu$K with the Boltzmann distribution as 
\small
\begin{equation}
K^\mathrm{el}(T) = \sqrt{\frac{8}{\pi \mu}} \left[ \frac{1}{k_BT}\right] ^\frac{3}{2} \int_{0}^{\infty}  E_\mathrm{c} \sigma^\mathrm{el}(E_\mathrm{c}) 
\mathrm{exp}\left(-\frac{E_\mathrm{c}}{k_BT}\right) dE_\mathrm{c},
\label{eq:Rate}
\end{equation}
\normalsize
where ${k_B}$ is the Boltzmann constant. 
We calculate $\sigma^\mathrm{el}(E_\mathrm{c})$ at nine collision energies of $E_\mathrm{c}=0.1, 0.5, 0.8, 1, 2, 3, 4, 6$, and  $8~\mu$K to cover the relevant region for integration in \cref{eq:Rate}. 
Numerical simulations in Ref.~\cite{ThielemannPRX25} suggest that, at a stray electric field of $E_\text{stray}\approx$ {mV/m} and a bath temperature of $T_\mathrm{Li}\approx  0.7~\mu$K, 
the collision energy distribution is well described by the Maxwell-Boltzmann distribution with  $T=0.8~\mu$K. 
We assign the peaks, whose intensities exceed 50\% of the maximum value of $K^\mathrm{el}$ in $B=240-340$~G, as resonances.

\vspace{0.1cm}
{\it{\bf Potential scaling}}

To determine the range of scaling factor $\lambda_S$ ($S=0, 1$), we calculate the scattering length $a_S$ with scaling the {\it ab initio} potential $V^\mathrm{ref}_S(R)$. 
On each scaled potential, we solve the 1-dimensional radial Schr\"odinger equation with the renormalized Numerov method at the collision energy of $E_\mathrm{c}=1$~nK as shown in  \cref{Fig_scatteringlength}.
From the peak-to-peak of $a_S$ including $\lambda_S=1$, 
we select the ranges as $0.9816\le \lambda_0 \le 1.0027$ and $0.9789 \le \lambda_1 \le 1.0094$ for generating a sample of singlet and triple potentials. 
The potentials are scaled only in the short-range part.
The continuity of potential energy curves is maintained using a switching function centered on $R = 40$~bohr, bridging the scaled short-range and the unscaled long-range \cite{SM}.

\begin{figure}[t!]
\begin{center}
\includegraphics[height=0.23
\textheight,keepaspectratio]{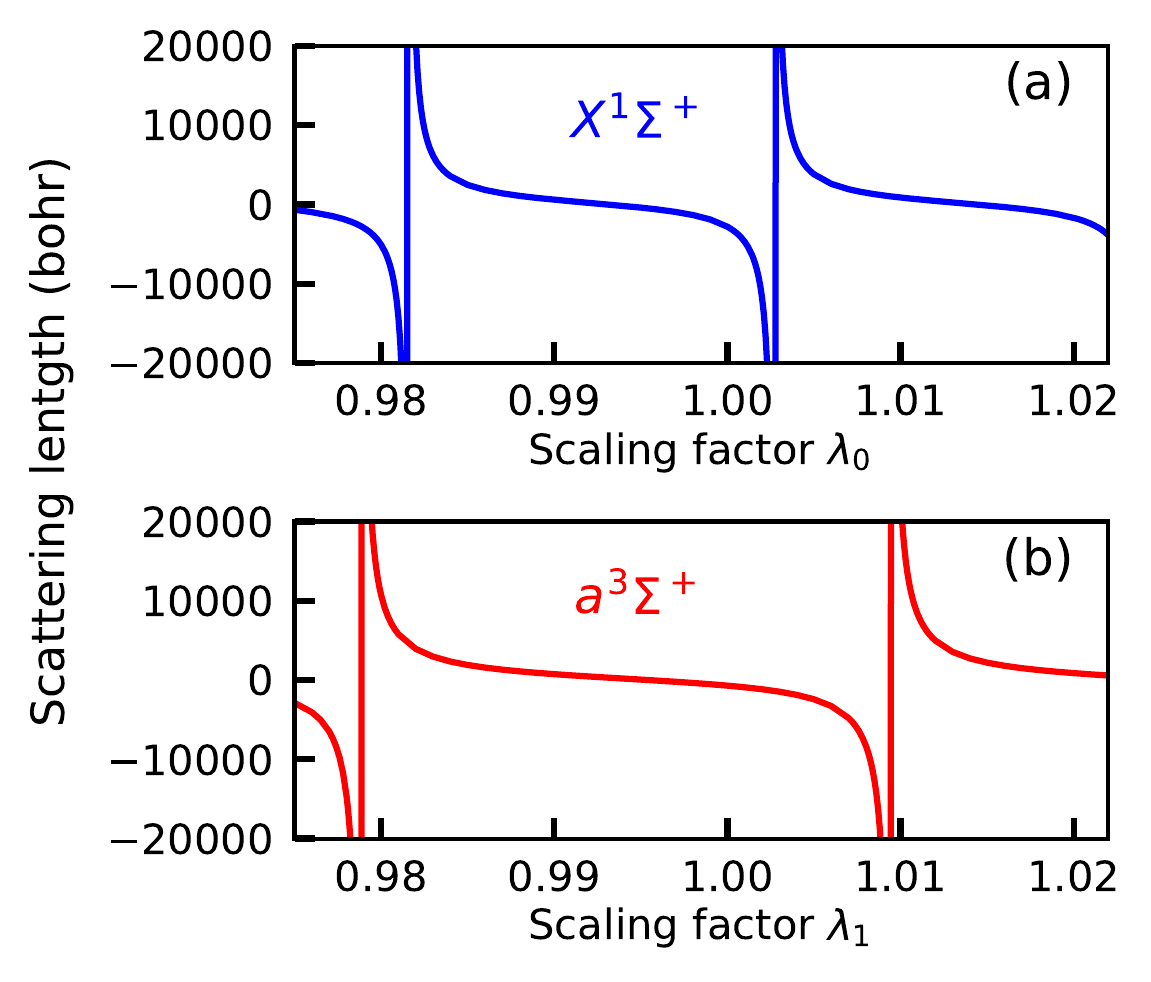}
\end{center}
\vspace*{-0.5cm}
\caption{Scattering length at $E_\mathrm{c}=1$~nK as a function of a scaling factor. (a) $V_0(R)$ ($X ^1\Sigma^+$) and (b) $V_1(R)$ $a ^3\Sigma^+$. 
}
\label{Fig_scatteringlength}
\end{figure}

\begin{figure}[t!]
 \centering
\includegraphics[width=0.91\columnwidth]{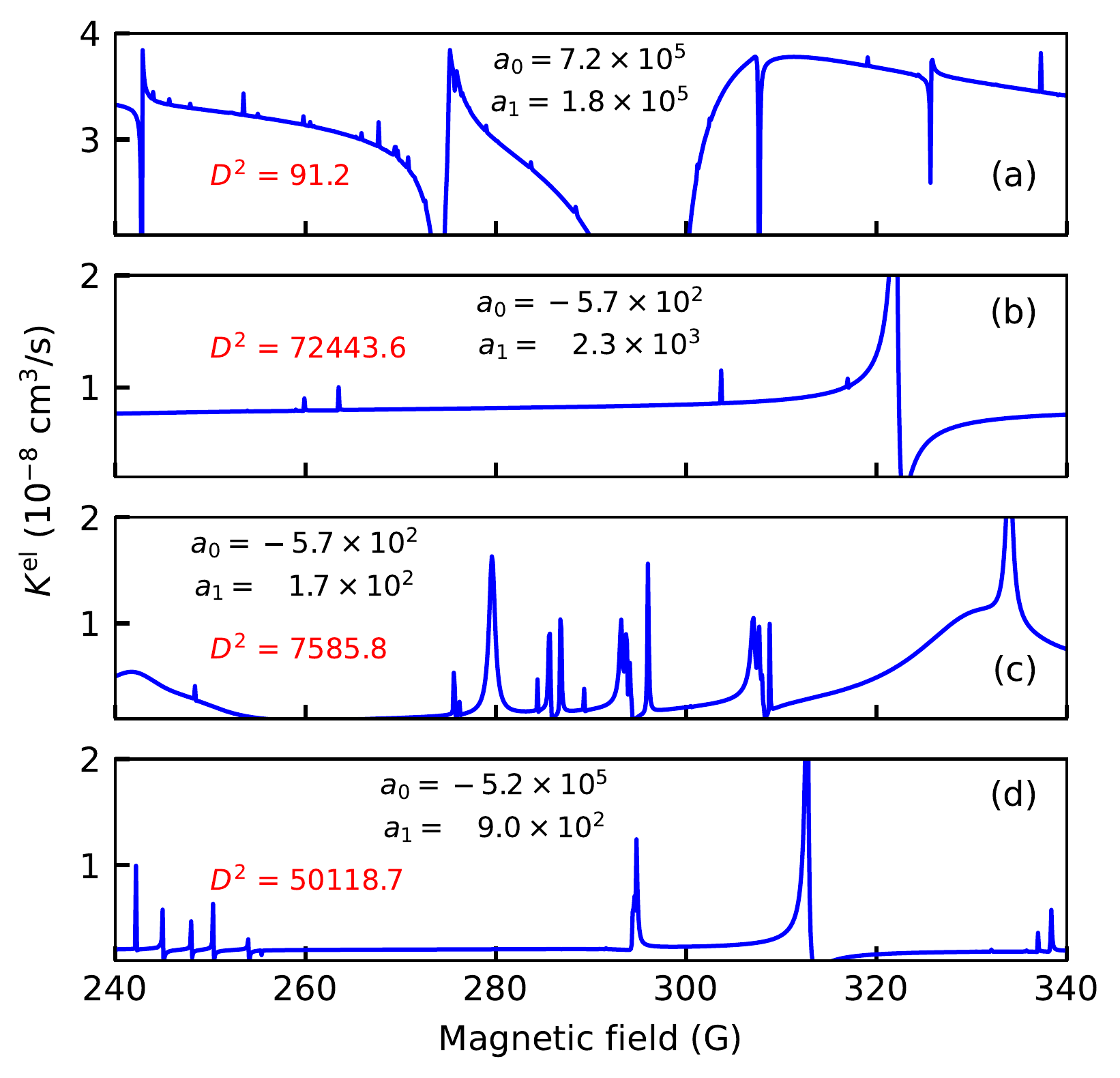}
    \vspace*{-0.5cm}
    \caption{ (a)-(d): Thermally averaged rate constant $K^\mathrm{el}(T)$ 
    at $T=0.8~\mu$K as a function of magnetic field
    with scaled potentials.
    $a_0$ and $a_1$ denote scaling factors for the singlet ($S=0$) and triplet ($S=1$) potentials.  
    The optimal potentials (a) minimize $D^2$ defined in \cref{eq:chi}.
    }
    \label{Fig_Rate}
\end{figure}

\Cref{Fig_Rate} shows examples of $K^\mathrm{el}$ calculated with scaled potentials, showing the sensitivity of the behavior of $K^\mathrm{el}$ to tiny changes in the potentials. 
\Cref{Fig_Rate} (a) shows the result with the optimal potentials taking the smallest value of $D^2$ (see \cite{SM} for more detailed information).

\vspace{0.1cm}
{\it{\bf Spin-orbit coupling}}

The SO interaction is taken into account in the Hamiltonian using the dipolar interaction form (\cref{eq:H_dipolar}) through $\lambda_\mathrm{SO}(R)$ determined based on the {\it ab initio} calculation \cite{TomzaPRA15}. For our CC calculations, we fit the {\it ab initio} $\lambda_\mathrm{SO}(R)$ to a simple analytic form as (in Hartree)
\small
\begingroup
\setlength\abovedisplayskip{5pt}
\setlength\belowdisplayskip{1.5pt}
\begin{equation}
\lambda_\mathrm{SO}(R)=0.091266\, \mathrm{exp}(-1.11R), 
\label{eq:scale_so}
\end{equation}
\normalsize
which describes \textit{ab initio} data well at intermediate and large distances~\cite{SM} (\cref{Fig_SO}).

Extensive exploration of the uncertainty in $\lambda_\mathrm{SO}(R)$ is beyond the scope of this Letter. We evaluated $D^2$ with three different scaled $\lambda_\mathrm{SO}(R)$, and found that the {\it ab initio} $\lambda_\mathrm{SO}(R)$ combined with the optimal potentials gives the minimum value of $D^2$ \cite{SM}. Thus, we use \cref{eq:scale_so} throughout the main text. 

\begin{figure}[b!]
\begin{center}
\includegraphics[height=0.14
\textheight,keepaspectratio]{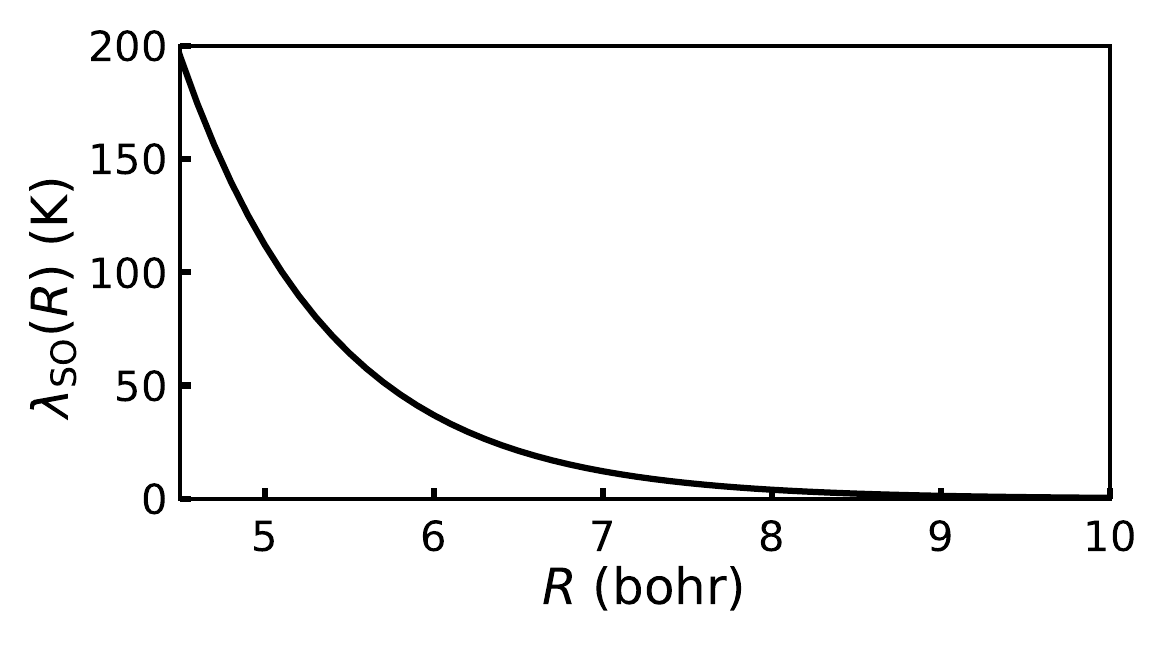}
\end{center}
\vspace*{-0.8cm}
\caption{ $R$-dependence of second-order spin-orbit coupling $\lambda_\mathrm{SO}(R)$.}
\label{Fig_SO}
\end{figure}

The effect of the SO coupling is demonstrated in \cref{Fig_Cross} using scaled potentials. We solve the CC equations, used in the scattering calculations, enforcing the bound state boundary condition \cite{HutsonCPC19_FIELD} to identify the bound states that cross the thresholds in $B=240-340$~G. 
Here, we assume the resolution limit of $\Delta B= 200$~{mG}, which is equal to the experimental resolution with $\ket{1}_\mathrm{Li}$, to distinguish individual resonances. 
Since overlaps of resonances arising from their intrinsic profiles as well as the collision energy distribution are not considered, the resultant number of resonances represents an upper bound on the number of resonances in $K^\mathrm{el}$ with $\Delta B= 200$~{mG}. 
For each threshold (each axis direction), we observe a significant change in the distribution of the number of resonances by including the SO interaction (ON) in the Hamiltonian, indicating the breaking of degeneracy (energy splitting) of bound states.   
We emphasize, regardless of potentials, that the number of resonances without SO (OFF) is much smaller than the experimental number of resonances (49) with $\ket{1}_\mathrm{Li}$.

\Cref{Fig_Cross} also shows the correlation between $\ket{1}_\mathrm{Li}$ and $\ket{2}_\mathrm{Li}$ in terms of the number of resonances. 
The moderate positive correlation ($r>0.6$) is consistent with a similar behavior of $\mathcal{N}(B)$ between $\ket{1}_\mathrm{Li}$ and $\ket{2}_\mathrm{Li}$ in \cref{Fig_1vs2} (see \cite{SM} for more details).

\begin{figure}[t!]
\begin{center}
\includegraphics[height=0.22
\textheight,keepaspectratio]{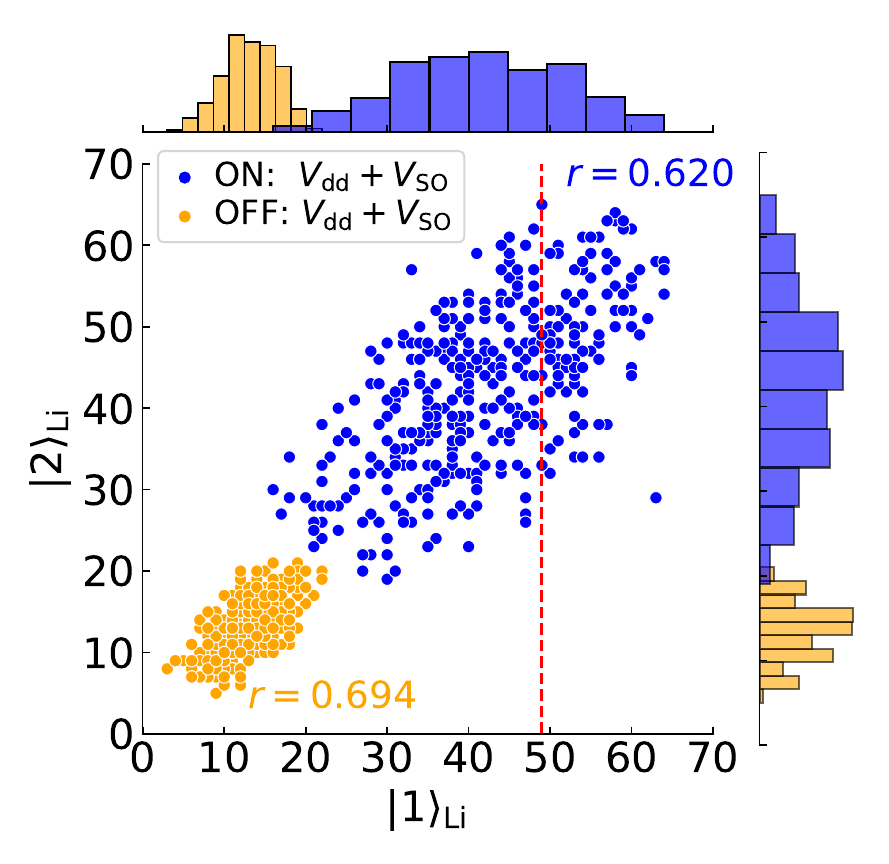}
\end{center}
\vspace*{-0.8cm}
    \caption{ Correlation between $\ket{1}_\mathrm{Li}$ and $\ket{2}_\mathrm{Li}$ for the number of resonances in $B=240-340$~G. 
    $r$: Pearson correlation coefficient. Red dashed line indicates the number of observed resonances in the experiment with $\ket{1}_\mathrm{Li}$. 
    }
\label{Fig_Cross}
\end{figure}

\vspace{0.12cm}
{\it{\bf Brody distribution}}

In this study, the unfolding is performed by mapping the resonance positions $B_i$ ($i=1,2,3,...$) onto the sequence defined as $\xi_i=\xi(B_i)$. Then the nearest neighbor spacing (NNS) is given as $s_i=\xi_{i+1}-\xi_i$ for $i=1,2,3,...$ \cite{FryePRA16,GreenPRA16,KosickiNJP20} as described in the main text.

Given a set of NNS \{$s_i$\}, the probability density function of it is expressed using the Brody distribution $P_\mathrm{B}(s;\eta)$ (\cref{eq:Brody}). The Brody parameter $\eta$ 
is determined by searching the value of $\eta$ that maximizes the likelihood function $\mathcal{L}(\eta)=\Pi_i  P_\mathrm{B}(s_i;\eta)$ or equivalently the log-likelihood function $\ell(\eta)=\sum_i \mathrm{ln} P_\mathrm{B}(s_i;\eta)$ 
\cite{SM,FryePRA16,GreenPRA16,KosickiNJP20}.

\Cref{Fig_PDF_Exp_vs_CC} shows the Brody distributions with scaled potentials. We observe a good agreement between the calculated result using the optimal potentials and the experimental one. The results with other potentials exhibit all types of behavior characterized by $\eta$ in $0\le\eta\le1$. 
This suggests that even small changes in the short-range part of the potential can affect not only individual resonances but also the statistical properties of resolvable resonances in the thermally averaged rate constant $K^\mathrm{el}$. 
We find the scaled potentials that reproduce the experimental Brody parameter ($\eta=0.25$). However, these potentials are not optimal in terms of $D^2$ defined in \cref{eq:chi}, and they produce significantly fewer resonances (27) than those observed experimentally (49).

\begin{figure}[h!]
\begin{center}
\includegraphics[height=0.14
\textheight,keepaspectratio]{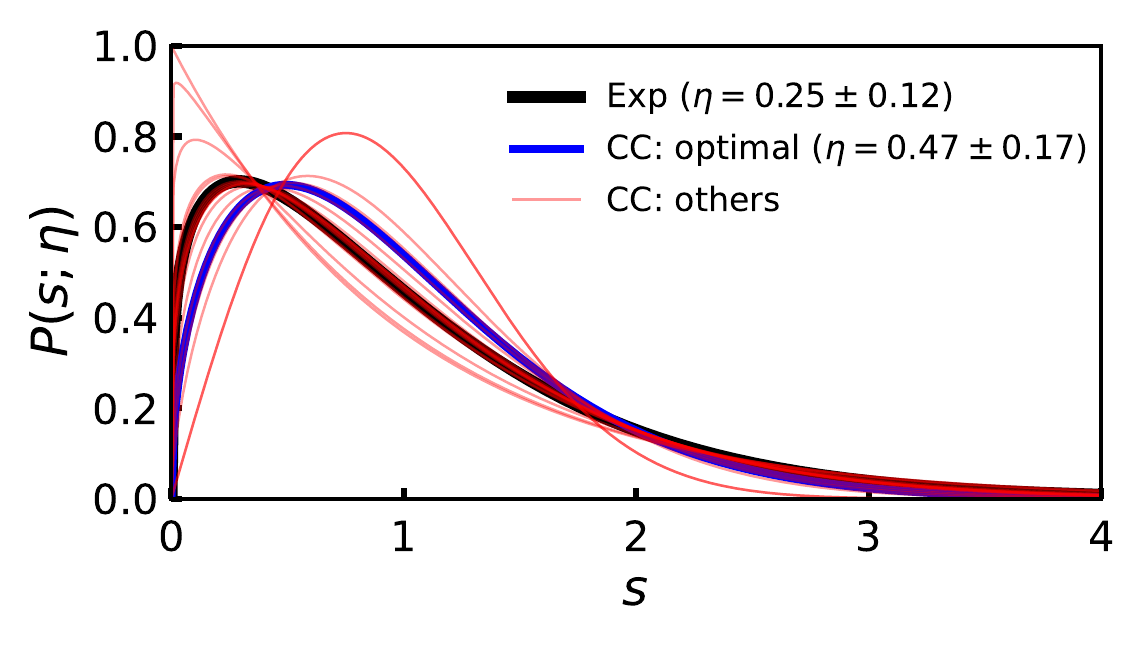}
\end{center}
\vspace*{-0.8cm}
\caption{Probability densities of
unfolded nearest neighbor spacing with a sample of scaled potentials and experimental spectrum for the initial state of $\ket{1}_\mathrm{Li}$. }
\label{Fig_PDF_Exp_vs_CC}
\end{figure}

\bibliography{Theory.bib}

\begin{thebibliography}{51}
\bibitem [{exp()}]{expfootnote}%
  \BibitemOpen
  \href@noop {} {}\bibinfo {note} {At higher temperatures we observe an increase of overall ion loss. This requires more ions to be reloaded and thus increases the average experiment cycle time. Hence, we limit the magnetic field scan to two sub-ranges in which we observe features already at the lowest collision energies.}\BibitemShut {Stop}%
\end{thebibliography}

\end{document}


\title{ \bf{Supplemental Material for \\  "Magnetic Feshbach resonances in Ba$^+$+Li collisions due to large spin-orbit coupling"}}

\author{Masato Morita}
\thanks{These authors contributed equally to this work.\\ \textcolor{blue}{masato.morita@fuw.edu.pl}}
\affiliation{Faculty of Physics, University of Warsaw, Pasteura 5, 02-093 Warsaw, Poland}

\author{Joachim Siemund}
\thanks{These authors contributed equally to this work.\\ \textcolor{blue}{masato.morita@fuw.edu.pl}}
\affiliation{Physikalisches Institut, Albert-Ludwigs-Universit\"at Freiburg, Hermann-Herder Str. 3, 79104  Freiburg, Germany}

\author{Wei Wu}
\affiliation{Physikalisches Institut, Albert-Ludwigs-Universit\"at Freiburg, Hermann-Herder Str. 3, 79104  Freiburg, Germany}

\author{Daniel von Schoenfeld}
\affiliation{Physikalisches Institut, Albert-Ludwigs-Universit\"at Freiburg, Hermann-Herder Str. 3, 79104  Freiburg, Germany}

\author{Jonathan Grieshaber}
\affiliation{Physikalisches Institut, Albert-Ludwigs-Universit\"at Freiburg, Hermann-Herder Str. 3, 79104  Freiburg, Germany}
    
\author{Agata Wojciechowska}
\affiliation{Faculty of Physics, University of Warsaw, Pasteura 5, 02-093 Warsaw, Poland}

\author{Krzysztof Jachymski}
\affiliation{Faculty of Physics, University of Warsaw, Pasteura 5, 02-093 Warsaw, Poland}

\author{Thomas Walker}
\affiliation{Physikalisches Institut, Albert-Ludwigs-Universit\"at Freiburg, Hermann-Herder Str. 3, 79104  Freiburg, Germany}
    \affiliation{Blackett Laboratory, Imperial College London, Prince Consort Road, London SW7 2AZ, United
    Kingdom}

\author{Fabian Thielemann}
\affiliation{Physikalisches Institut, Albert-Ludwigs-Universit\"at Freiburg, Hermann-Herder Str. 3, 79104  Freiburg, Germany}
\affiliation{5. Physikalisches Institut and Center for Integrated Quantum Science and Technology,
Universität Stuttgart, Pfaffenwaldring 57, 70569 Stuttgart, Germany}

\author{Tobias Schaetz}
\email{tobias.schaetz@physik.uni-freiburg.de}
\affiliation{Physikalisches Institut, Albert-Ludwigs-Universit\"at Freiburg, Hermann-Herder Str. 3, 79104  Freiburg, Germany}

\author{Micha{\l} Tomza}
\email{michal.tomza@fuw.edu.pl}
\affiliation{Faculty of Physics, University of Warsaw, Pasteura 5, 02-093 Warsaw, Poland}

\maketitle
\tableofcontents

\vspace{0.5cm}
In this Supplemental Material (SM), we provide details of the technical information as well as additional results to complement the main text. In \cref{sec:SM_hyperfine-Zeeman},
we present fundamental information about the hyperfine-Zeeman states of Li and identify the most surprising aspect in the experimental spectra, explaining the reason.
In \cref{sec:SM_scale1D}, we describe some technical details for the potential scaling method employed in the main text. We demonstrate the necessity of taking into account the degeneracy breaking of bound states, due to the spin-orbit (SO) coupling, to obtain a reasonable number of resonances (\cref{sec:SM_SO}). We also discuss the positive correlation between $\ket{1}_\mathrm{Li}$ and $\ket{2}_\mathrm{Li}$ in terms of the number of resonances. 
We provide an expression used for the SO parameter $\lambda_\mathrm{SO}(R)$. 
The sum of the squared minimum distances $D^2$ is calculated in \cref{sec:SM_scaling_chi} using various scaled potentials and SO parameters.  The associated staircase functions $\mathcal{N}(B)$ are presented in \cref{sec:SM_scaling_stair}. The behavior of the likelihood function with unfolded nearest neighbor spacing (NNS) and the uncertainty of the determined Brody parameter are discussed in \cref{sec:SM_likelihood}. The spectral rigidity is shown in \cref{sec:SM_rigidity} to explore the long-range correlation of the resonances.
The effects of the coexistence of different initial states, $\ket{1}_\mathrm{Li}$ and $\ket{2}_\mathrm{Li}$, are explored in \cref{sec:SM_Li2}. 
We show the temperature dependence of the elastic rate constant $K^\mathrm{el}(T)$ in \cref{sec:SM_Temp}.

\setcounter{equation}{0}
\setcounter{figure}{0}

\newpage

\section{\label{sec:SM_hyperfine-Zeeman} Hyperfine-Zeeman state}

The major subject in the present Letter is the surprising initial Li state dependence of the number and distribution of resonances in the experimental spectra. To clarify which point is the most surprising, we briefly summarize the magnetic field ($B$-field) dependence of the character of the lowest $\ket{1}_\mathrm{Li}$ and the second lowest $\ket{2}_\mathrm{Li}$ hyperfine-Zeeman states of $^6$Li. 
\newline

Near $B=0$~G,  $\ket{1}_\mathrm{Li}$ and  $\ket{2}_\mathrm{Li}$ are characterized by the hyperfine states $\ket{F=1/2, M_F=+1/2}$ and $\ket{ F=1/2, M_F=-1/2 }$ 
\begin{equation}
\begin{aligned}
\ket{1}_\mathrm{Li}=\ket{F=1/2, M_F=+1/2} &=\;\;\; \sqrt{2/3}\, \ket{M_S=-1/2}  \ket{M_I=+1}
+ \sqrt{1/3}\, \ket{M_S=+1/2}\ket{M_I=\;\;\;0}, 
\\
\ket{2}_\mathrm{Li}=\ket{ F=1/2, M_F=-1/2 } &= -\sqrt{1/3}\, \ket{M_S=-1/2}\ket{M_I=\;\;\;0} 
+ \sqrt{2/3}\, \ket{M_S=+1/2}  \ket{M_I=-1}.
\label{eq:hyperfineZeeman0G}
\end{aligned}
\end{equation}
The expressions with direct product states are shown in the right-hand side of \cref{eq:hyperfineZeeman0G}, indicating that the down spin state $\ket{M_S=-1/2}$ and the up spin state $\ket{M_S=+1/2}$ are comparable in both $\ket{1}_\mathrm{Li}$ and $\ket{2}_\mathrm{Li}$. The weights of the $\ket{M_S=-1/2}$ and $\ket{M_S=+1/2}$ states are 2/3 and 1/3 for $\ket{1}_\mathrm{Li}$ and 1/3 and 2/3 for $\ket{2}_\mathrm{Li}$.  
\newline

On the other hand, $\ket{1}_\mathrm{Li}$ and $\ket{2}_\mathrm{Li}$ converge to $\ket{M_S=-1/2}\ket{M_I= 1}$ and $\ket{M_S=-1/2}\ket{M_I= 0}$, respectively, with increasing $B$-field due to the dominance of the electron Zeeman interaction that decouples the hyperfine coupling as
\begin{equation}
\begin{aligned}
\ket{1}_\mathrm{Li}\approx\ket{M_S=-1/2}  \ket{M_I=+1}&=\;\,\;\sqrt{2/3}\ket{F=1/2,M_F=+1/2} - \sqrt{1/3}\ket{F=3/2,M_F=+1/2},
\\
\ket{2}_\mathrm{Li}\approx\ket{M_S=-1/2}  \ket{M_I=\,\,\,\:0}&=-\sqrt{1/3}\ket{F=1/2,M_F=-1/2} + \sqrt{2/3}\ket{F=3/2,M_F=-1/2}.
\end{aligned}
\end{equation}
These show that the electron spin in both $\ket{1}_\mathrm{Li}$ and $\ket{2}_\mathrm{Li}$ is characterized by the down spin state $\ket{M_S=-1/2}$. We note that the collision-induced electron spin-exchange with $^{138}$Ba$^+$ does not occur at ultracold energies for these states due to the hyperfine interaction for $^6$Li. While the electron spin-exchange between the upper Zeeman state of Ba$^+$ and $\ket{2}_\mathrm{Li}=\ket{M_S=-1/2}  \ket{M_I=\,\,\,\:0}$ can occur if $\ket{2}_\mathrm{Li}$ is excited to $\ket{6}_\mathrm{Li}=\ket{M_S=+1/2}  \ket{M_I=\,\,\,\:0}$, such excitation is not energetically allowed because the excitation energy of it is larger than the magnitude of the de-excitation energy of the upper Zeeman state of Ba$^+$. The excitation from $\ket{1}_\mathrm{Li}=\ket{M_S=-1/2}  \ket{M_I=+1}$ to $\ket{6}_\mathrm{Li}=\ket{M_S=+1/2}  \ket{M_I=+1}$ requires a larger energy.
\newline

The present $B$-field range ($B=240-340$~G) corresponds to the latter case. With $B=240$~G, $\ket{1}_\mathrm{Li}$ and $\ket{2}_\mathrm{Li}$ are given as
\begin{equation}
\begin{aligned}
\ket{1}_\mathrm{Li}&=0.990\ket{M_S=-1/2}  \ket{M_I=+1} -0.139 \ket{M_S=+1/2}  \ket{M_I=\,\,\,\:0}, 
\\
\ket{2}_\mathrm{Li}&=0.985\ket{M_S=-1/2}  \ket{M_I=\,\,\,\:0}-0.172\ket{M_S=+1/2}  \ket{M_I=-1}. 
\label{eq:hyperfineZeeman240G}
\end{aligned}
\end{equation}
Therefore, the weights for the down spin state $\ket{M_S=-1/2}$ for $\ket{1}_\mathrm{Li}$ and $\ket{2}_\mathrm{Li}$ are 0.98 and 0.97, respectively. \Cref{eq:hyperfineZeeman240G} clearly shows that both these states are characterized by the down spin state $\ket{M_S=-1/2}$ and the energies for these states exhibit a quite similar $B$-field dependence as shown in Fig.~1 (b) in the main text. Although the major nuclear spin states are different between $\ket{1}_\mathrm{Li}$ and $\ket{2}_\mathrm{Li}$ as $\ket{M_I=+1}$ and $\ket{M_I=\,\,\,\:0}$, there is no nuclear spin-dependent interaction in the Hamiltonian other than the hyperfine interaction for Li. Therefore, the difference in the character of the nuclear spin states does not directly cause the difference in the electron spin dynamics. 

As long as we consider collisions in the framework of simple (yet well-established) two-body collision theory, there is no clear basis to expect a strong dependence on these initial states. Therefore, the observed differences between $\ket{1}_\mathrm{Li}$ and $\ket{2}_\mathrm{Li}$, both in terms of individual resonances and statistical properties, are unexpected and puzzling.

\section{\label{sec:SM_scale1D} Potential scaling}
\vspace{-0.2cm}
The electron spin-exchange interaction between Ba$^+$($^2$S) and Li($^2$S) is characterized by the difference of singlet $V_0(R)$ and triplet $V_1(R)$ potentials in the short-range. 
In turn, the long-range parts of $V_0(R)$ and $V_1(R)$  are described by a common long-range force, which guarantees the short-range character of the spin-exchange interaction \cite{SikorskyPRL18}. 
In this study, we use a common long-range force constant, $C_4 = 80.91$~a.u. The long-range part, given in the form of $-C_4/R$, is not scaled in generating a sample of potentials. 
To maintain the continuity of the potential energy curves, a switching function centered on $R=40$~bohr bridges the short-range part and the long-range part. 
The employed switching function is expressed in the form of $\mathrm{sw}(R)=(1/2)\{1-\mathrm{tanh}(R-40)\}$. 
The scaled potential is written as $V^\mathrm{scaled}_S(R) = \mathrm{sw}(R)\times \lambda_{S} V_S^\mathrm{ref}(R)+(1-\mathrm{sw}(R))\times (-C_4/R^4),\, (S=0,1)$, where $V_S^\mathrm{ref}(R)$ denotes the {\it ab initio} potentials.

\vspace{-0.3cm}
\section{\label{sec:SM_SO} Spin-orbit interaction}
\vspace{-0.2cm}

\begin{figure}[b!]
\begin{center}
\includegraphics[height=0.25
\textheight,keepaspectratio]{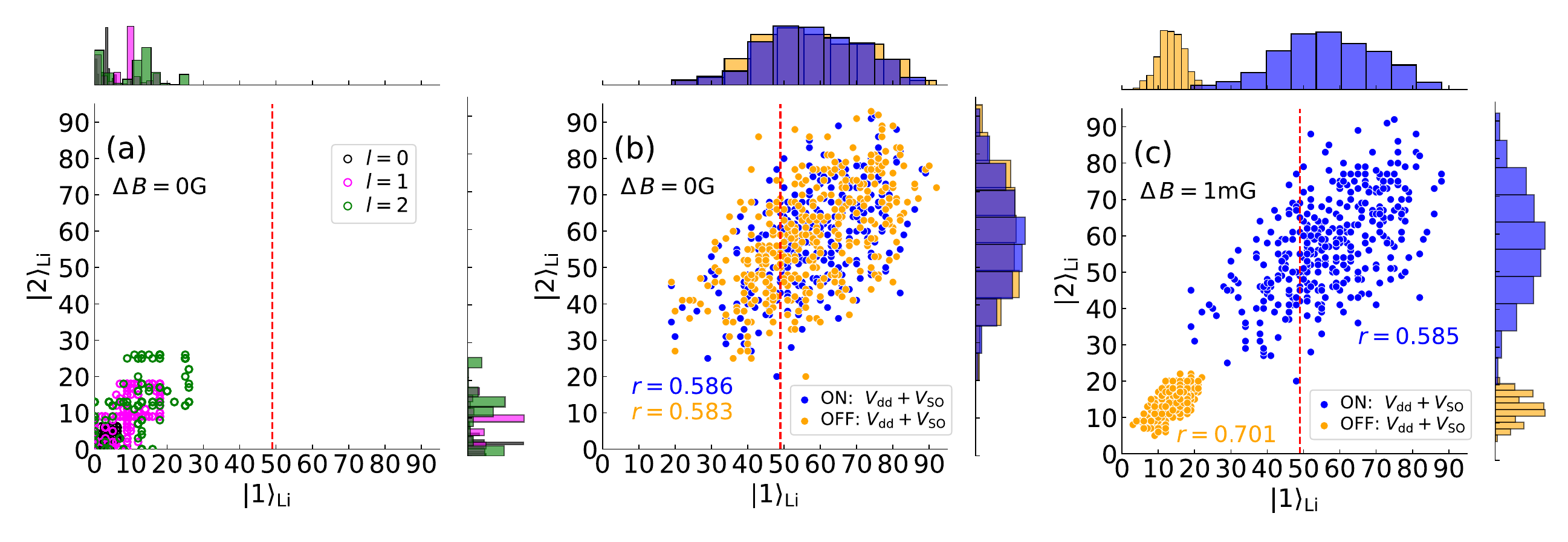}
\end{center}
\vspace*{-0.8cm}
\caption{ Correlation between $\ket{1}_\mathrm{Li}$ and $\ket{2}_\mathrm{Li}$ in terms of the numbers of resonances in $B=240-340$~G using a sample of scaled potentials. (a) Number of resonances for each partial wave component without SO, (b) total number of resonances (c) total number of resonances with the resolution limit of $\Delta B=1$~mG. $r$: Pearson’s product-moment correlation coefficient.
Red dashed line: number of resonances (49) observed in the experimental loss spectrum with $\ket{1}_\mathrm{Li}$.
}
\label{Fig_Cross}
\end{figure}

To clarify the influence of the spin-orbit (SO) coupling on the number of resonances, we perform bound state calculations with (ON) and without (OFF) the SO interaction (\cref{Fig_Cross}) using a sample of scaled potentials (the sample size is 400). We solve the CC equations with the boundary condition for bound state \cite{HutsonCPC19_FIELD} and count the number of bound states whose energies cross the thresholds of $\ket{1}_\mathrm{Li}$ and $\ket{2}_\mathrm{Li}$ in the range of $B=240-340$~G. 
These crossings may cause resonances in the rate constants $K^\mathrm{el}$. As mentioned in End Matter of the main text, the number of crossings provides the upper bound of the number of resonances in $K^\mathrm{el}$ obtained with the same basis set and potentials. 
Throughout this section, we simply refer to the upper bound of the number of resonances as "the number of resonances".
The correlations between $\ket{1}_\mathrm{Li}$ and $\ket{2}_\mathrm{Li}$ in terms of the number of resonances are shown as scatter plots. 

We show the results for each partial wave component $l=0-2$ without SO and magnetic dipole-dipole interactions in \cref{Fig_Cross} (a). We note that each partial wave component is independent because of the absence of SO. The number of resonances in the panel is counted assuming infinite resolution in the magnetic field ($\Delta B=0$~G) and also resolving exactly degenerated bound states.  
The number of resonances for each component is quite small. As a result, the total number of resonances with $l=0-2$ is less than the number of resonances 49 (vertical red dashed line) observed experimentally. Therefore, we can confirm that treating only $l=0-2$ is not sufficient to reproduce the experimental number of resonances. 
As described in the main text, we employ six partial wave components, $l=0-5$, in the basis set for the scattering calculations to obtain $K^\mathrm{el}$. We use the same basis set for the panels \cref{Fig_Cross} (b) and (c). 

In \cref{Fig_Cross} (b), we consider the distributions of the total number of resonances. We note that the SO interaction induces the coupling between different partial waves, thus the decomposition into each independent partial wave component is usually not possible for the number of resonances. 
The panel (b) shows similar resonance number distributions between the ON and OFF cases, for both thresholds, which is reasonable because, as a whole, the SO coupling does not generate new bound states. Furthermore, the correlation between $\ket{1}_\mathrm{Li}$ and $\ket{2}_\mathrm{Li}$ is also similar between ON and OFF. The moderate positive correlations ($r\approx0.58$) are consistent with a similarity of $\mathcal{N}(B)$ between $\ket{1}_\mathrm{Li}$ and $\ket{2}_\mathrm{Li}$ in Fig.~4 of the main text.

As we count all the resonances regardless of overlap or degeneracy with other resonances in (b), we do not find a significant difference in the resonance number distribution between the presence (ON) and absence (OFF) of the SO interaction. However, the SO coupling can largely change the distribution of the resonances. This change is highlighted if we introduce a resolution limit between individual resonances as shown in (c). 
Without the SO coupling, there are many degenerate resonances with a common partial wave $l$. If we assume a finite resolution limit in B-field ($\Delta B$) to resolve individual resonances, we observe a significant decrease in the number of resonances without SO (OFF).  In other words, the SO coupling increases the number of resolvable resonances, breaking the degeneracy of resonances.  

\begin{figure}[t!]
\begin{center}
\includegraphics[height=0.19
\textheight,keepaspectratio]{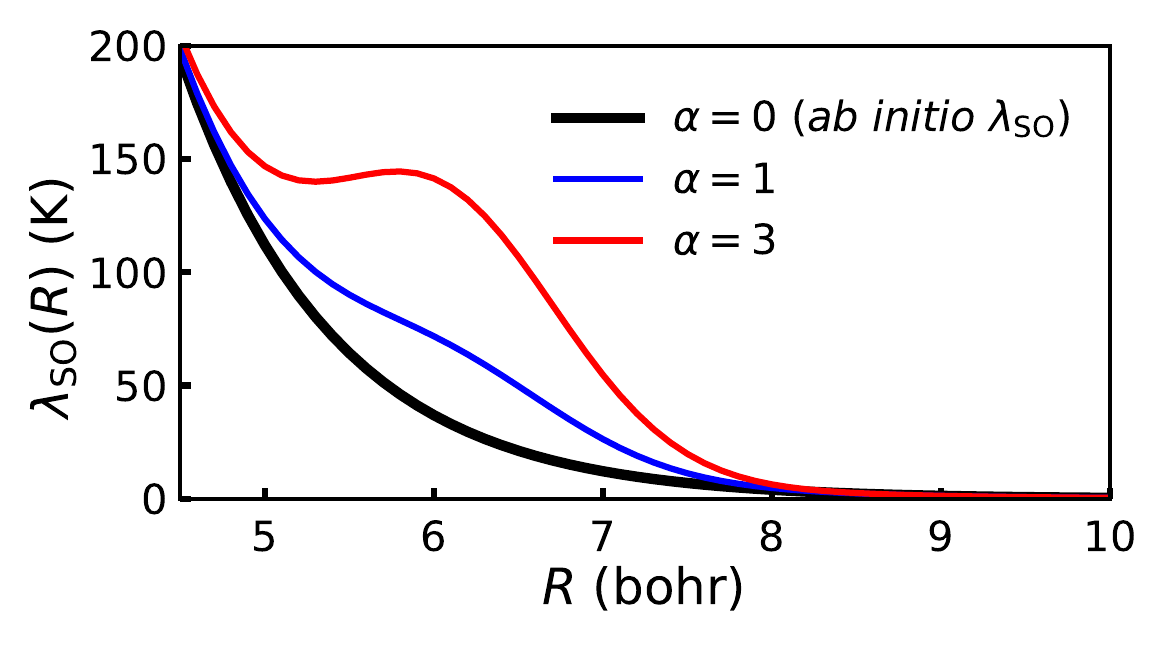}
\end{center}
\vspace*{-0.8cm}
\caption{Second-order spin-orbit coupling parameter $\lambda_\mathrm{SO}(R)$ in kelvin as a function of $R$. $\alpha$ denotes the strength of the Gaussian term in \cref{eq:scale_so}.}
\label{Fig_SO}
\end{figure}

The difficulty of resolving individual resonances due to the overlap of resonances may also arise from the widths and magnitudes of resonances and the collision energy distribution. 
Therefore, it is necessary to integrate the degeneracy breaking and the overlapping of resonances as we perform in the main text by calculating the thermally averaged $K^\mathrm{el}$.
However, from \cref{Fig_Cross} (c), we can emphasize that the number of resonances without SO (OFF) is much smaller than the experimental number of resonances (49) with $\ket{1}_\mathrm{Li}$ regardless of potentials.   
We note that the number of resonances with $\ket{2}_\mathrm{Li}$ might be slightly overestimated due to the presence of artificial {\it box-quantized} continuum states lying above the threshold of $\ket{1}_\mathrm{Li}$. We also note that the SO interaction can generate additional resonances. As mentioned above, the SO interaction typically does not create bound states. However, it introduces additional couplings between the bound and continuum states via the mixing of relative rotation motion ($l,M_l$) and electron spins ($M_{S_\mathrm{Ba^+}}, M_{S_\mathrm{Li}}$).

Although we exclusively show the results obtained with the {\it ab initio} $\lambda_\mathrm{SO}(R)$ in the main text, we explore the effect of the change in $\lambda_\mathrm{SO}(R)$ and present the results in the following chapters, \cref{sec:SM_scaling_chi,sec:SM_scaling_stair}. We express the deviation from the original {\it ab initio} $\lambda_\mathrm{SO}(R)$ using a Gaussian function as  
\begin{equation}
\lambda_\mathrm{SO}(R)=0.091266\, \mathrm{exp}(-1.11R)+\alpha(1.1061\times 10^{-4})\,\mathrm{exp}[-(R-6.05)^2],
\label{eq:scale_so}
\end{equation}
where the first term shows the behavior of {\it ab initio} $\lambda_\mathrm{SO}(R)$ as described in End Matter of the main text. The center of the second Gaussian term, $R=6.05$~bohr, corresponds to the distance at which $^3\Sigma^+$ crosses $b^3\Pi$, the prefactor $1.1061\times10^{-4}$ comes from the value of the 1st term at $R=6.05$~bohr, namely  $0.091266\, \mathrm{exp}(-1.11\times6.05)$.  Therefore,  $\alpha=0$ in the second term results in the original {\it ab initio} $\lambda_\mathrm{SO}(R)$ and $\alpha=1$ doubles the height of $\lambda_\mathrm{SO}(R)$ at $R=6.05$~bohr (see \cref{Fig_SO}). 

The dipolar form of $V_\mathrm{SO}(R)$ (Eq.~(2) in the main text) implies that $\lambda_\mathrm{SO}(R)$ is derived using perturbation theory for the electronic states. However, the perturbative approach breaks down when the energy difference between the $a^3\Sigma^+$ and $b^3\Pi$ electronic states vanishes at the crossing of the potentials as shown in Fig.~1 (c) of the main text. Indeed, the SO interaction can significantly enhance or suppress the effective coupling strength, particularly around the crossing at $R \approx 6$~bohr. Beyond the perturbative approach, the effective SO coupling depends on the energy separation between the collision threshold and vibrational levels on the $^3\Pi$ potential (see the inset of Fig.~1 (c)), which unfortunately cannot be predicted accurately enough from the calculated potential, thus leaving the exact strength of the second-order SO coupling as a free parameter of our model. However, the crossing between the $a^3\Sigma^+$ and $b^3\Pi$ states occurs at an energy region far below the scattering thresholds at a very short distance, making this coupling a short-range effect. 

\vspace{0.1cm}
\section{\label{sec:SM_scaling_chi} $D^2$ with scaled potentials }
\vspace{0.1cm}
\begin{figure}[b!]
\begin{center}
\includegraphics[height=0.28
\textheight,keepaspectratio]{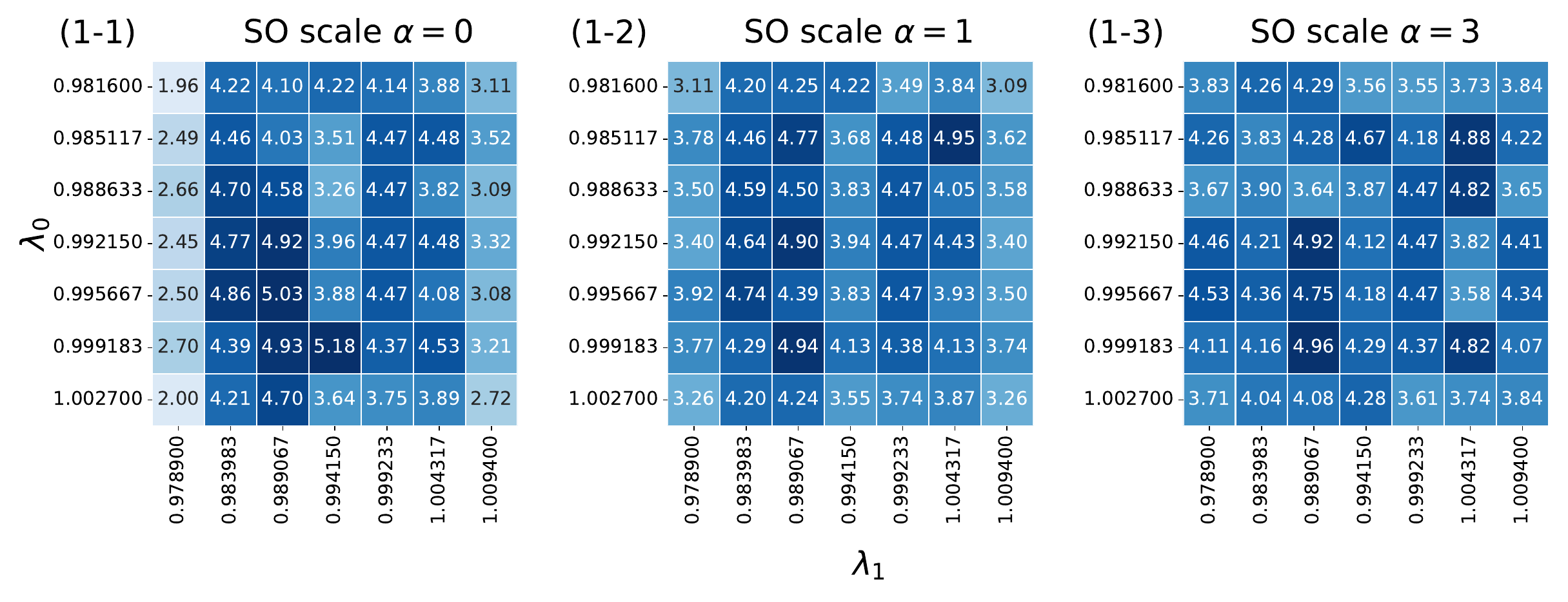}
\end{center}
\begin{center}
\includegraphics[height=0.28
\textheight,keepaspectratio]{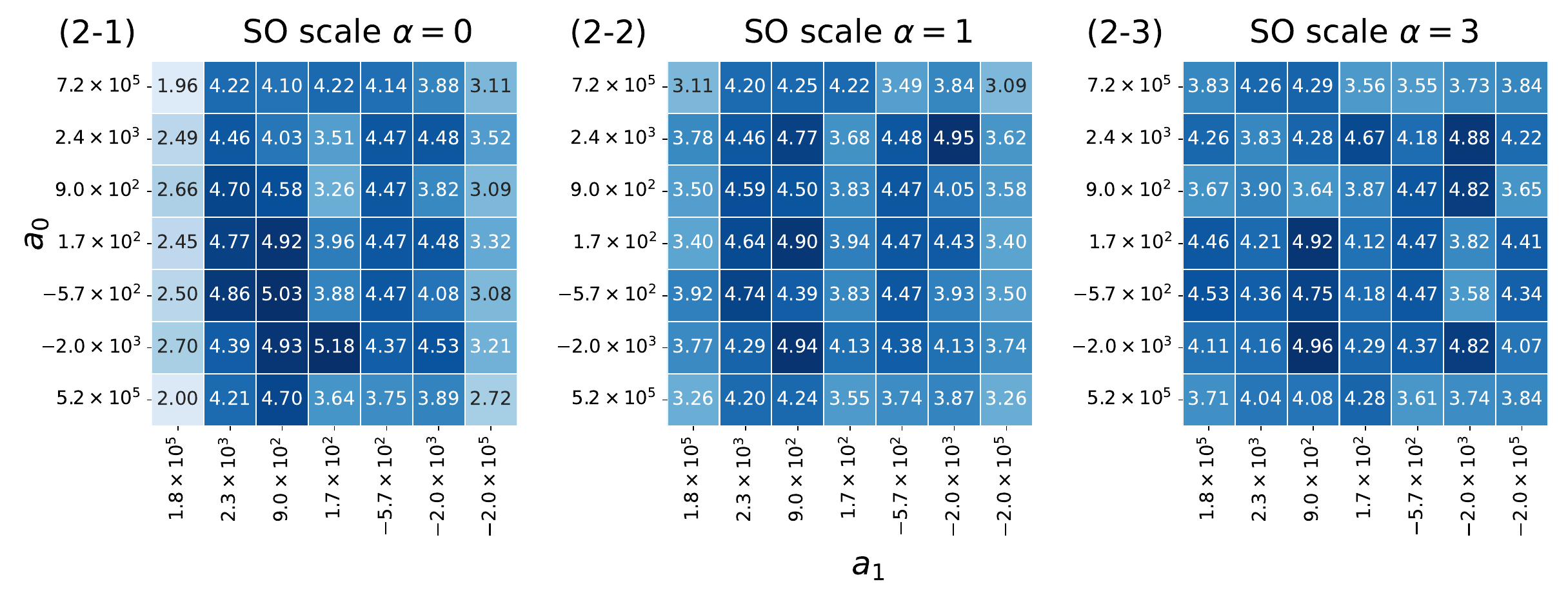}
\end{center}
\caption{ (1-1)-(1-3): $\mathrm{log_{10}}D^2$ with potentials obtained by multiplying scaling factors $\lambda_{S} ($S=0,1$)$ to the {\it ab initio} singlet ($S=0$) and triplet ($S=1$) potential energy curves, $V_0^\mathrm{ref}(R)$ and $V_1^\mathrm{ref}(R)$. A factor of $\alpha$, changing the {\it ab inito} spin-orbit parameter $\lambda_\mathrm{SO}(R)$ as shown in \cref{eq:scale_so}, is set as 0 (1-1), 1 (1-2) and 3 (1-3). 
Panels (2-1)-(2-3) are identical to the upper ones while the axes with $\lambda_S$ are replaced by those with corresponding scattering lengths $a_S$  (see End Matter of the main text).}
\label{Fig_log10chi}
\end{figure}

In the main text, we propose using $D^2$, the sum of squared distances from each experimental resonance to its nearest neighbor resonance in $K^\mathrm{el}$,
as an objective function to optimize the singlet ($S=0$) and triplet ($S=1$) potentials. 
We systematically change the value of the scaling factors $\lambda_S, ($S=0,1$)$ in the ranges of $0.9816\le \lambda_0 \le 1.0027$ and $0.9789 \le \lambda_1 \le 1.0094$, then obtain a sample of the scaled potentials.
With the initial state of $\ket{1}_\mathrm{Li}$, we calculate the value of $D^2$ based on the resonance positions of the experimental loss spectrum and the calculated rate constant $K^\mathrm{el}$ 
in $B=240-340$~G. In the main text, the peaks in $K^\mathrm{el}$, whose heights are larger than half of the maximum value of $K^\mathrm{el}$ in $B=240-340$~G, are detected as resonances. 
Although the number of resonances may increase if we relax the criterion for the peak height, $D^2$ and $\mathcal{N}(B)$, calculated using the optimal potentials, are stable against the criterion.

Complete exploration in the three-dimensional parameter space of the scaling factors, $\lambda_0$, $\lambda_1$, and $\alpha$ in \cref{eq:scale_so} requires enormous computational costs. Therefore, we explore $\alpha$ with three different values of 0, 1, and 3. The upper panels of \cref{Fig_log10chi} show the values of $\mathrm{log_{10}}D^2$ with respect to the combinations of $\lambda_0$ and $\lambda_1$. The lower panels are identical to the upper ones, but the values of $\lambda_0$ and $\lambda_1$ are replaced with the corresponding scattering lengths (see End Matter in the main text). The optimal factors, which provide the minimum value of $D^2$, are identified as $\lambda_0=0.9816$, $\lambda_1=0.9789$, and $\alpha=0$. Thus, we focus on the calculations with $\alpha=0$, corresponding to the original {\it ab initio} $\lambda_\mathrm{SO}(R)$, in the main text.

\section{\label{sec:SM_scaling_stair} Staircase function $\mathcal{N}(B)$ }
\vspace{0.2cm}
\begin{figure}[b!]
\begin{center}
\includegraphics[height=0.48
\textheight,keepaspectratio]{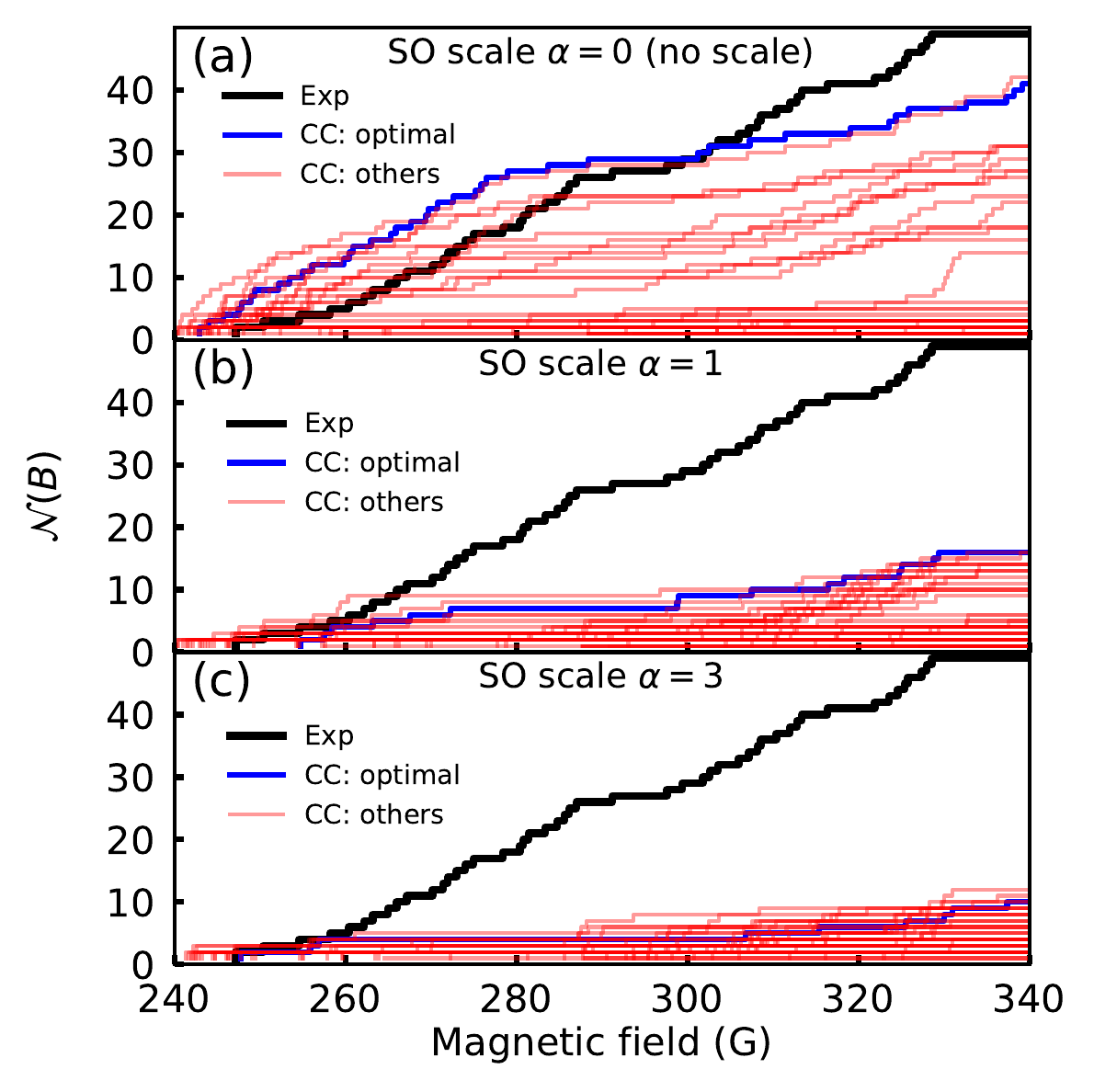}
\end{center}
\caption{Staircase functions for the resonances obtained with the experimental loss spectrum and the rate constants $K^\mathrm{el}$ with scaled potentials. Optimal potentials are given by the scaling factors of ($\lambda_0$, $\lambda_1$) = ($0.9816$, $0.9789$), corresponding to ($a_0$, $a_1$)=($7.152\times 10^5$, $1.838\times 10^5$).
}
\label{Fig_stair_SO}
\end{figure}

The staircase functions are shown in \cref{Fig_stair_SO}, calculated with the scaled potentials. Note that (a) shows the results with {\it ab initio} $\lambda_\mathrm{SO}(R)$ ($\alpha=0$ in \cref{eq:scale_so}) and is also shown in the main text. 
With increasing $\alpha$, we observed an overall trend of decreasing the number of resonances. At the same time, the number of resonances decreases from the results with $\alpha=0$ if we omit the SO interaction. Therefore, the {\it ab initio} $\lambda_\mathrm{SO}(R)$ provides a better description of the staircase functions than others.  

Since the SO interaction with a typical strength does not cause a change in the number of bound states, the observed change in the number of resonances mainly comes from the shift of the peaks accompanying the changes of the coupling strengths, which can both increase the number of detectable resonances by lifting degeneracies and decrease it by causing overlaps with other pronounced or degenerate peaks.

\section{\label{sec:SM_likelihood} Likelihood function }
\vspace{-0.1cm}

\begin{figure}[t!]
 \centering
 \includegraphics[width=0.8\columnwidth]{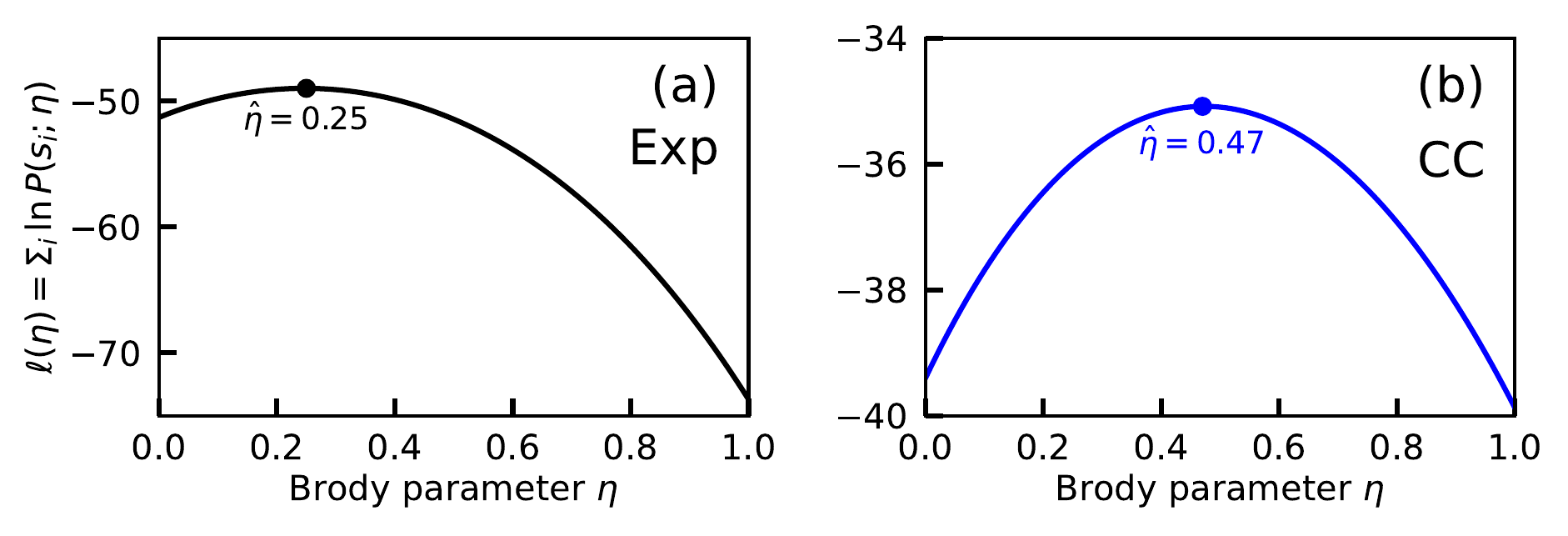}
    \caption{ Log-likelihood as a function of Brody parameter. (a) experimental loss spectrum and (b) $K^\mathrm{el}$ obtained by the CC calculation using the potentials optimized in the main text.}
    \label{Fig_likelihood}
\end{figure}

We calculate the Brody parameter $\eta$ based on the unfolded NNS distribution of resonances, \{$s_i$\} ($i=1,..,N$), using the maximum likelihood estimation. 
The maximum likelihood estimate (MLE) $\hat{\eta}$ for the Brody distribution $P_\mathrm{B}(s;\eta)$ is determined by finding the value of $\eta$ that maximizes the likelihood function $\mathcal{L}(\eta)=\Pi_i  P_\mathrm{B}(s_i;\eta)$ or equivalently the log-likelihood function $\ell(\eta)=\sum_i \mathrm{ln} P_\mathrm{B}(s_i;\eta)$. We show the behavior of $\ell(\eta)$ as a function of $\eta$ in \cref{Fig_likelihood} for the experimental loss spectrum and $K^\mathrm{el}$ obtained by the CC calculation with the optimal potentials for $\ket{1}_\mathrm{Li}$.  

The standard deviation of the maximum likelihood estimate, as an indicator of the statistical uncertainty of the estimated parameter, is given as $\sigma\approx\sqrt{1/N\mathcal{I}(\eta_0)}$, where $\eta_0$ is the true parameter and $\mathcal{I}(\eta_0)$ is the Fisher information at $\eta_0$ defined by the expectation as $\mathcal{I}(\eta_0) = \mathbb{E} [ ( \partial \ln P_\mathrm{B}(s; \eta)/\partial \eta )^2 ]_{\eta=\eta_0} = -\mathbb{E} [ ( \partial^2 \ln P_\mathrm{B}(s; \eta)/\partial \eta^2) ]_{\eta=\eta_0}$ \cite{CasellaBOOK02}.
Note that the Fisher information $\mathcal{I}(\eta)$ is the variance of the score function $U(\eta;s) = \partial \ln L(\eta; s)/{\partial \eta}$ with respect to $s$, where the likelihood function $L$ is given as $L(\eta; s)=P_\mathrm{B}(s; \eta)$. 
We numerically approximate $\mathcal{I}(\eta_0)$ by replacing the unknown true parameter $\eta_0$ with the MLE $\hat{\eta}$ obtained with \{$s_i$\}. We use $\pm \sigma$ as the 68\% confidence interval for the estimated Brody parameter as shown in Fig.~3 (b) of the main text (Exp: $\sigma=0.1400$, CC: $\sigma=0.1784$). 
In practice, the total Fisher information $N\mathcal{I}(\eta_0)$ is often approximated simply by the observed Fisher information, which is the second derivative of the negative log-likelihood function $l(\eta)$ (negative curvature) evaluated at MLE (see \cref{Fig_likelihood}), therefore
$\sigma\approx\sqrt{(-d^2\ell (\eta)/d\eta^2)_{\eta=\hat{\eta}}^{-1}}$ \cite{GreenPRA16,FryePRA16}. We confirmed that this approximation provides similar results (Exp: $\sigma=0.1539$, CC: $\sigma=0.1908$).

\section{\label{sec:SM_rigidity} Spectral rigidity }
\vspace{-0.1cm}

Spectral rigidity is used to assess the long-range correlation of the unfolded spectrum. Dyson and Mehta introduced the spectral rigidity $\Delta_3(\Delta \xi)$ as the minimum mean square deviation between the staircase function \(N(\xi)\) and the best linear approximation \(A \xi + B\) within a spectral interval $[\xi_{\mathrm{0}}, \xi_{\mathrm{0}} + \Delta \xi]$ \cite{DysonJMP63}. 
Similarly to the number variance $\Sigma^2$ shown in the main text, it is averaged over the starting points of the interval $\xi_0$ for numerical stability \cite{BohigasPRL84, HaqPRL82, GuhrPR98}. We evaluate the (averaged) spectral rigidity as a function of the length of the interval $\Delta \xi$, 
\begin{equation}
\Delta_3(\Delta \xi) = \left\langle \min_{A,B} \frac{1}{\Delta \xi} \int_{\xi_{\mathrm{0}}}^{\xi_{\mathrm{0}} + \Delta \xi} \bigl( N(\xi) - A \xi - B \bigr)^2 \, d\xi \right\rangle_{\xi_{\mathrm{0}}},
\label{eq:rigidity}
\end{equation}
where the slope $A$ and the intercept $B$ are determined by minimizing the squared deviation in the least squares sense. The averaging \(\langle \cdot \rangle_{\xi_{\mathrm{0}}}\) is taken over all possible starting points of the interval \(\xi_{\mathrm{0}}\) within the unfolded spectrum.

\begin{figure}[t!]
 \centering
 \includegraphics[width=0.4\columnwidth]{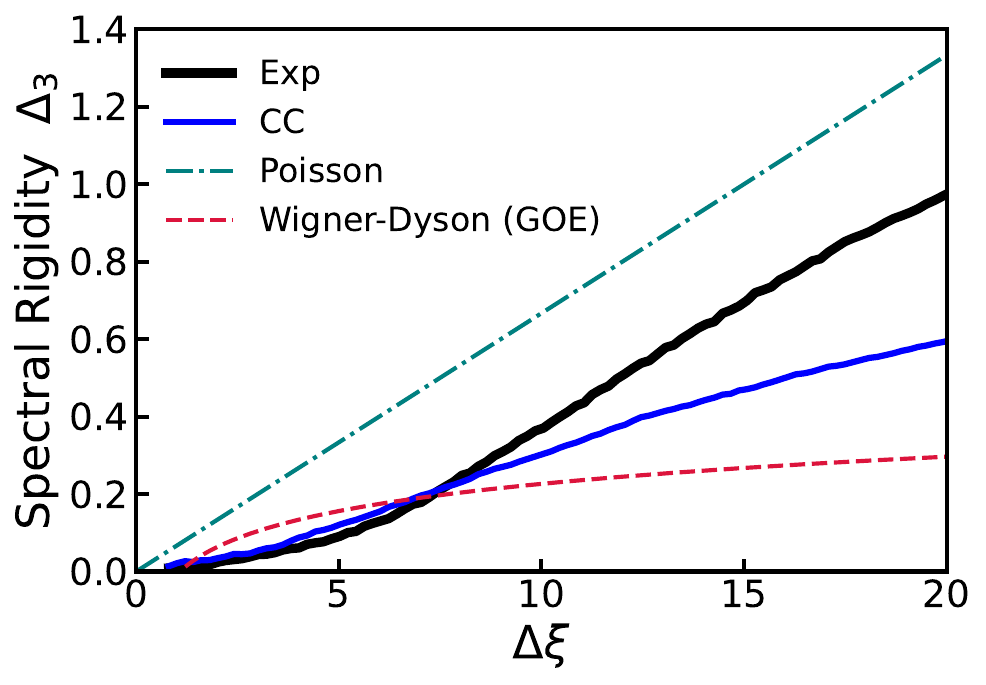}
    \caption{Spectral rigidity as a function of the length of interval $\Delta \xi$. }
    \label{Fig_rigidity}
\end{figure}

\Cref{Fig_rigidity} shows the spectral rigidity for the experimental loss spectrum (Exp) and $K^\mathrm{el}$ (CC) calculated with the optimal potentials with the initial state of $\ket{1}_\mathrm{Li}$. Although the theoretical result (CC) agrees well with the experimental result (Exp) in the short interval length $\Delta \xi < 8$, their behavior deviates from Poisson or Gaussian orthogonal ensemble (GOE). Since there is only a small number of resonances in an interval of very short length $\Delta Xi$ (sometimes no resonances), the result with small $\Delta \xi$ tends to be unstable or largely influenced by outliers. With longer intervals, the trend of the results is very similar to that obtained with $\Sigma^2$ in the main text. Both the experimental and theoretical results behave as intermediate cases between the Poissonian and GOE cases, and the experimental result is more similar to the Poissonian case than the theoretical one. \\

\section{\label{sec:SM_Li2} Collision with distinct initial states }

Our optimization of potentials is carried out based on the collisions between Ba$^+$ and Li in their lowest states. 
Although we found a significant discrepancy between the experiment and the calculation in terms of resonance density with the initial state of $\ket{2}_\mathrm{Li}$, a slight discrepancy in the staircase functions is observed even with $\ket{1}_\mathrm{Li}$.

\begin{figure}[h!]
\centering
\includegraphics[width=0.62\columnwidth]{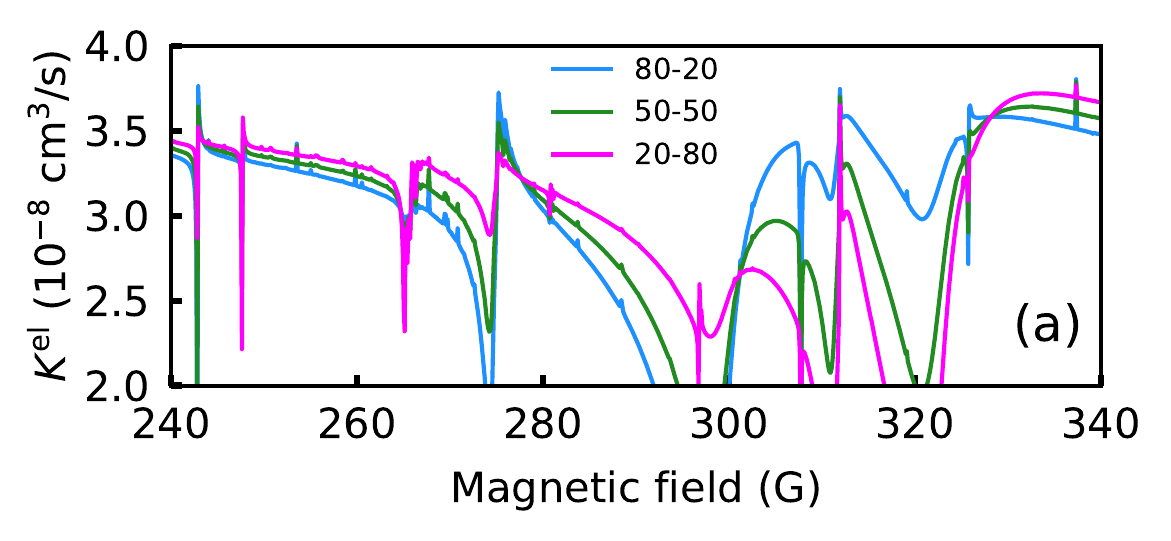}
\includegraphics[width=0.62\columnwidth]{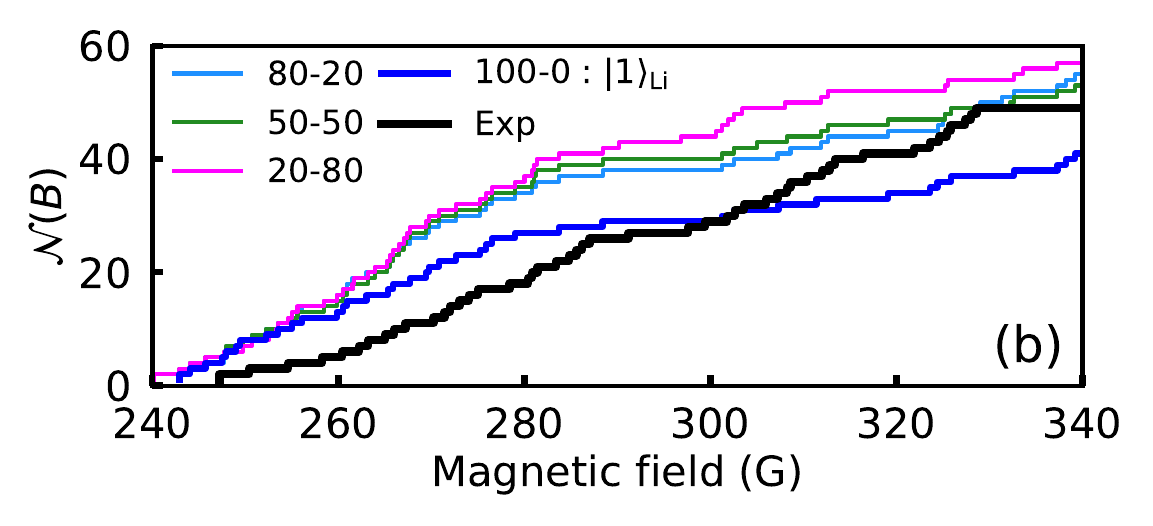}

    \caption{ (a) Rate constant as a function of magnetic field with three different abundance ratio between $\ket{1}_\mathrm{Li}$ and $\ket{2}_\mathrm{Li}$ for the initial state of Li. $\ket{1}_\mathrm{Li}$ : $\ket{2}_\mathrm{Li}$ = $ 80\%\,:\,20\%$ (light blue), $50\%\,:\,50\%$ (light green), and $20\%\,:\,80\%$ (magenta). CC:$\ket{1}_\mathrm{Li}$  (blue) corresponds to 100-0 ($100\%\,:\,0\%$).
    The calculations are performed with the optimal potentials in the $D^2$ analysis as discussed in the main text and \cref{sec:SM_scaling_chi}.
    (b) Staircase function $\mathcal{N}(B)$.  }
    \label{Fig_1mix2}
\end{figure}

One of the sources of the discrepancies might be unexpected contamination of different initial states. Here, we consider the contamination of $\ket{2}_\mathrm{Li}$. 
Due to the difference of resonance positions between $\ket{1}_\mathrm{Li}$ and $\ket{2}_\mathrm{Li}$ as shown in the main text, the mixing of $\ket{1}_\mathrm{Li}$ and $\ket{2}_\mathrm{Li}$ can affect the resultant staircase function as shown in the lower panel of \cref{Fig_1mix2}. It should be emphasized that a slight contamination of $\ket{2}_\mathrm{Li}$ leads to a large difference in the resultant number of resonances in $K^\mathrm{el}$. See light blue (20\% contamination of $\ket{2}_\mathrm{Li}$) and blue (pure $\ket{1}_\mathrm{Li}$).

\vspace{0.7cm}
\section{\label{sec:SM_Temp} Temperature dependence of the rate constant }

In this Letter, the thermally averaged rate constants $K^\mathrm{el}(T)$ are calculated assuming $T=0.8~\mu$K for the Maxwell-Boltzmann distribution, based on the simulation discussed in Ref.~\cite{ThielemannPRX25}. 

\begin{figure}[h!]
 \centering
    \includegraphics[width=0.6\columnwidth]{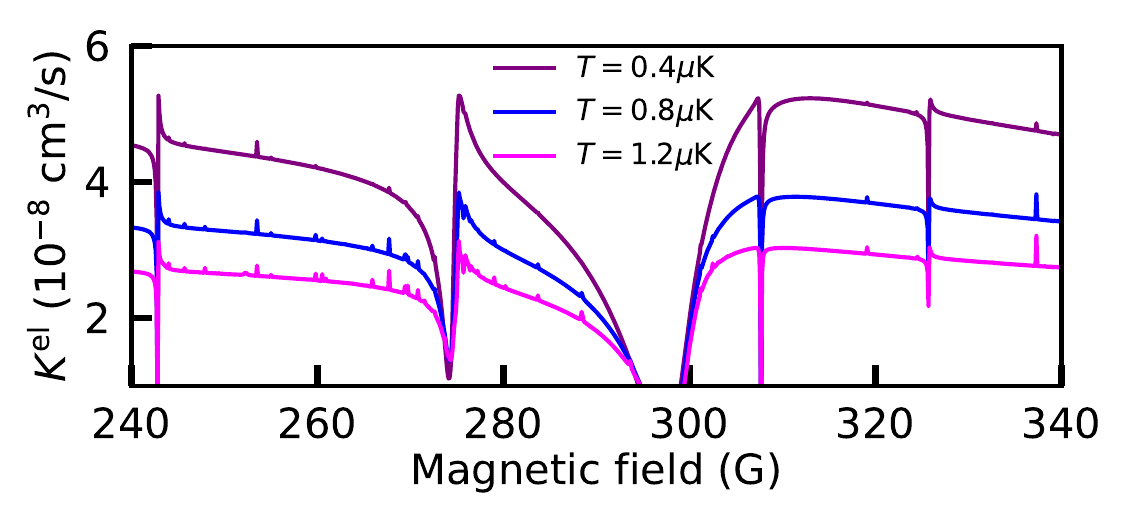}
    \caption{ Rate constant at the temperatures of $T=0.4, 0.8$, and $1.2\mu$K.  }
    \label{Fig_Rate_temp}
\end{figure}

Finally, we examine the sensitivity of $K^\mathrm{el}(T)$ to the temperature. 
In \cref{Fig_Rate_temp}, we calculate $K^\mathrm{el}(T)$ at three different temperatures for the collisions with $\ket{1}_\mathrm{Li}$. The resonance positions are stable concerning temperature. We observe that the relative importance of small resonance peaks increases with temperature, which is consistent with the relative importance of higher partial waves at higher temperatures. However, we note that the decrease of the overall (background) magnitude with increasing temperature would also play a role in emphasizing the visibility of the minor resonance peaks in the figure.

\newpage
\bibliography{Theory.bib}